\documentclass[%
 reprint,
 amsmath,amssymb,
 aps,
prb,
floatfix,
]{revtex4-2}
\usepackage{graphicx}
\usepackage{dcolumn}
\usepackage{bm}
\usepackage{hyperref}

\usepackage{xcolor}
\usepackage{makecell}

\begin{document}

\preprint{APS/123-QED}

\title{ Room-temperature optomechanics with light-matter condensates}

\author{Vladislav~Yu.~Shishkov\textsuperscript{1,2}}

\author{Evgeny~S.~Andrianov\textsuperscript{1,2}}

\author{Anton~V.~Zasedatelev\textsuperscript{3}}
\email{anton.zasedatelev@univie.ac.at}

\affiliation{
 \textsuperscript{1}Dukhov Research Institute of Automatics (VNIIA), 
 22~Sushchevskaya, Moscow~127055, Russia
}
\affiliation{
 \textsuperscript{2}Moscow Institute of Physics and Technology, 9~Institutskiy~pereulok, Dolgoprudny~141700, Moscow~region, Russia
}

\affiliation{
\textsuperscript{3}Vienna Center for Quantum Science and Technology~(VCQ),
~Faculty~of~Physics,~University~of~Vienna, Boltzmanngasse~5, 1090~Vienna, Austria
}

\date{\today}

\begin{abstract}
In this work, we develop an optomechanical formalism for macroscopic quantum states in exciton-polariton systems with strong exciton-phonon interactions. We show that polariton optomechanical interactions induce dynamical backaction, resulting in dispersive and dissipative shifts in the complex vibrational response functions. Unlike conventional optomechanical systems, polariton optomechanics features high-dimensionality and phase-space confinement due to the dispersion relations of exciton-polaritons. Consequently, vibrational modes exhibit effective positive or negative mass depending on the detuning parameter, and are capable for the nonequilibrium vibrational Bose-Einstein condensation under the resonant conditions~\cite{shishkov2023mapping}. We demonstrate the potential for vibrational control of polariton condensates at room temperature.

\end{abstract}

\maketitle


\section{Introduction}

Cavity optomechanics with molecules has recently emerged as a new research frontier promising quantum control over the vibrational states of matter coupled to optical cavities at room temperature ~\cite{roelli2016molecular,schmidt2017linking,esteban2022molecular}. High frequencies of molecular vibrations on the order of a few tens of THz ensure quantum ground state preparation of the mechanical subsystem even at high temperatures~\cite{esteban2022molecular,kasperczyk2016temporal}. An ultra-small mode volume of plasmonic cavities enables strong optomechanical interaction with single molecules~\cite{benz2016single}, achieving the coupling strength of 20 meV ($g\sim5$ THz). Mostly implemented within surface-enhanced Raman scattering (SERS) configuration, molecular optomechanics holds the record high single-photon optomechanical interaction strength among existing platforms~\cite{esteban2022molecular}. Besides fundamental interest, such strong optomechanical interaction is practical for energy transduction, bridging mid-IR and visible spectral ranges with applications in ultrafast photon detection operational at room temperature~\cite{chen2021continuous,xomalis2021detecting}. However, the extreme optical mode localization in plasmonic cavities results in significant losses ($\kappa\sim 25$ THz), which prevent the observation of strong single-photon optomechanical interactions ($g / \kappa > 1$) in current experiments. Additionally, high cavity losses necessitate high pumping rates to develop coherent vibrational phases via parametric instabilities and significantly constrain dynamical backaction rates to the Doppler regime for vibrational modes below a cut-off frequency ($\omega_{\rm Vib} < \kappa/2$)~\cite{aspelmeyer2014cavity}. While sideband-resolved optomechanical interactions can be implemented for sufficiently high vibrational resonances, their amplification toward a coherent state is unlikely due to the potential breakdown of molecular bonds at such high vibrational occupations~\cite{lombardi2018pulsed}.

\begin{figure}
\includegraphics[width=1\linewidth]{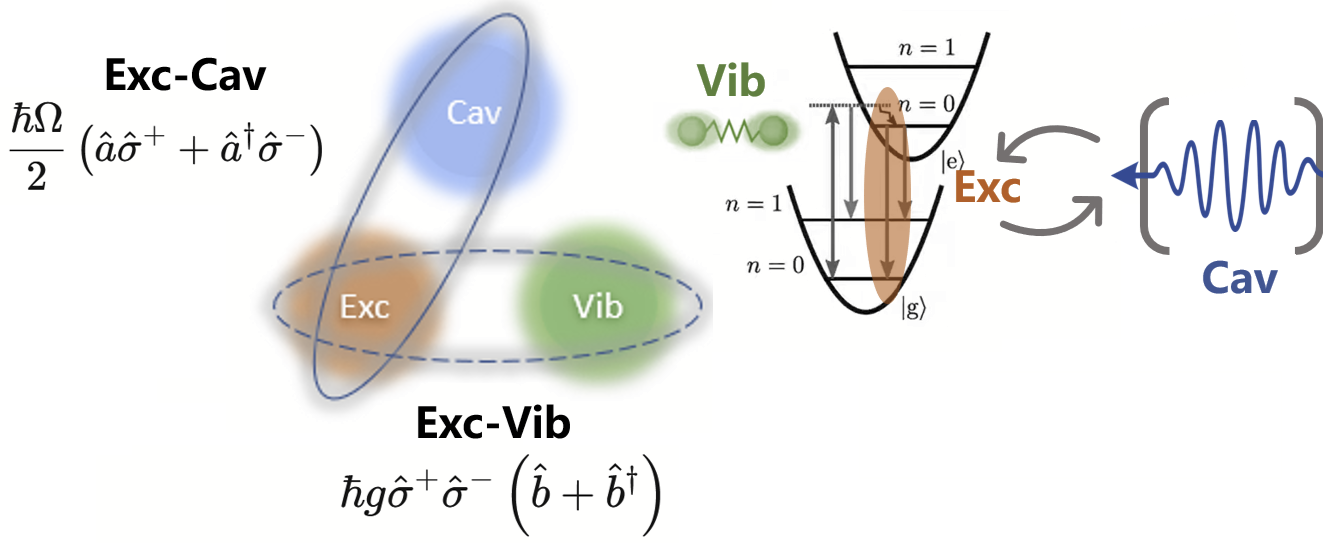}
\caption{
Schematic of polariton optomechanics based on the tripartite strong interaction between excitons, vibrations, and cavity photons.
}
    \label{fig:Fig1}
\end{figure}

In this work, we explore a new regime in molecular quantum optomechanics characterized by: (i) - collective coupling of large molecular ensembles of $\sim 10^{8}-10^{12}$ molecules, within high-Q microcavities; and (ii) - simultaneous strong exciton-photon and exciton-vibration interactions. Fig.~\ref{fig:Fig1} illustrates the tripartite interaction involved in the optomechanical coupling. It exhibits a resonant polariton nature due to the strong exciton-photon interaction and low losses $\omega_{\rm Vib} \gg \kappa,\gamma$ providing conditions for the well-resolved sideband regime as well as strong optomechanical backaction to lift the limits on parametric control over vibrational states. Finally, under the resonant blue-detuned laser drive we enter a macroscopic quantum state of polariton Bose--Einstein condensation accompanied by the build-up of coherent vibrational phases. In Ref.~\cite{shishkov2023mapping}, we propose a new sympathetic mechanism to achieve vibrational BEC via genuine polariton optomechanics. 




The article is organized as follows: in Section~\ref{sec:TheoreticalFramework}, we present the theoretical framework introducing Hamiltonian, commutation relations, interaction with the environment, and the equations of motion for the collective and localized excitonic and vibrational states. 
In Section~\ref{sec:susceptibility}, we develop a coherent optomechanical framework for the vibrational mechanism of polariton condensation. We introduce an effective optomechanical Hamiltonian that leads to dynamical backaction effects analogous to those in a cavity with a suspended mirror under radiation pressure in the resolved-sideband regime~\cite{aspelmeyer2014cavity}.
In Section~\ref{sec:consequences}, we present numerical simulations of vibrational and polariton occupation numbers as functions of optomechanical detuning and laser drive power. In the resonant blue-detuned optomechanical configuration, we observe strong vibrational amplification leading to condensation of molecular vibrations into the lowest energy state. We propose non-resonant methods for parametric vibrational control over polariton Bose--Einstein condensates, compatible with telecom wavelengths.

\section{Theoretical framework} \label{sec:TheoreticalFramework}

We explore a nonequilibrium microscopic model representing an ensemble of molecules with vibrationally dressed electronic transitions, coupled to cavity modes characterized by in-plane momenta $\hbar {\bf k}_{\parallel}$ (hereinafter $\hbar{\bf k}$). Fig.~\ref{fig:Fig1} sketches this setup, depicting vibrationally dressed excitons coupled to an optical cavity mode, which leads to tripartite effective polariton optomechanical interaction.
Originally suggested for strongly coupled GaAs microcavities, polariton optomechanical interaction so far involves delocalized Wannier-Mott type excitons and low-frequency phonon states~\cite{kyriienko2014optomechanics, jusserand2015polariton, carlon2022enhanced, santos2023polaromechanics, kuznetsov2023microcavity}. 
In this work, electronic excitations being localized at molecules are effectively treated as Frenkel-type excitons, where each excitation can be described by the Pauli creation and annihilation operators acting on a single molecule. 
We take into account a vibrational state with eigenfrequency $\omega_{\rm Vib}$ in each molecule.
The system Hamiltonian has the form
\begin{multline}\label{Full Hamiltonian simple form}
\hat H =  
\sum\limits_{\bf k} 
\hbar {\omega_{{\rm Cav}|{\bf k}}}
\hat a_{{\rm Cav}|{\bf k}}^\dag {{\hat a}_{{\rm Cav}|{\bf k}}}
+
\\
\sum\limits_{j=1}^{N_{\rm mol}} 
{\hbar {\omega_{{\rm exc}}}\hat \sigma_{{\rm Exc}{j}}^\dag {{\hat \sigma}_{{\rm Exc}{j}}}}
+
\sum\limits_{j=1}^{N_{\rm mol}} 
{\hbar {\omega_{\rm Vib}}\hat b_{{\rm Vib}j}^\dag {\hat b_{{\rm Vib}j}}}
+
\\
\sum\limits_{j=1}^{N_{\rm mol}} 
{\hbar \Lambda \omega_{\rm Vib}\hat \sigma_{{\rm Exc}j}^\dag {{\hat \sigma }_{{\rm Exc} {j}}}\left( {{\hat b_{{\rm Vib}j}} + \hat b_{{\rm Vib}j}^\dag } \right)}
+
\\
\sum\limits_{j=1}^{N_{\rm mol}} 
\sum\limits_{{\bf k}} 
\hbar \Omega_{j{\bf k}}
\left( 
\hat \sigma_{{\rm Exc}{j}}^\dag \hat a_{{\rm Cav}|{\bf k}} e^{i{\bf k}{\bf r}_j}
+
h.c.
\right)
,
\end{multline}
where $\hat a_{{\rm{Cav}}|{\bf{k}}}^\dag$ (${\hat a_{{\rm{Cav}}|{\bf{k}}}}$) is the creation (annihilation) operator of a photon in the cavity with the wavevector ${\bf{k}}$. 
The corresponding eigenfrequency of the cavity photon is ${\omega _{{\rm{Cav}}|{\bf{k}}}}$. 
The commutation relation $ \left[ \hat a_{{\rm Cav}|{\bf k}},\hat a_{{\rm Cav}|{\bf k'}}^\dag \right] = \delta _{{\bf k},{\bf k'}}$ holds for the operators of the cavity,
$N_{\rm mol}$ is the total number of molecules in the illuminated region.
We assume, that each molecule can host one exciton with the frequency ${\omega _{{\rm{exc}}}}$  independently on the state of nearby molecules~\cite{yamamoto2003semiconductor}.
For $j$-th molecule $\hat \sigma_{{\rm{Exc}}{j}}^\dag$ (${\hat \sigma_{{\rm{Exc}}{j}}}$) is the creation (annihilation) operator of the exciton. 
Exciton operators obey anti-commutation relation ${{\hat \sigma}_{{\rm{Exc}}{j}}} {{\hat \sigma}_{{\rm{Exc}}{j}}^\dag} + {{\hat \sigma}_{{\rm{Exc}}{j}}^\dag} {{\hat \sigma}_{{\rm{Exc}}{j}}}  = 1$. 
Below we consider the case of small probability for the exciton to be found in an excited state, $\langle \hat \sigma^\dag_{{\rm{Exc}}j} \hat \sigma_{{\rm{Exc}}j} \rangle \ll 1 $. 
In this case, the approximate commutation relation $\left[ {{{\hat \sigma}_{{\rm{Exc}}{j}}},\hat \sigma_{{\rm{Exc}}{j'}}^\dag } \right] \approx {\delta _{{j},{j'}}}$ is valid~\cite{combescot2008microscopic}. 
We use rotating wave approximation in Hamiltonian Eq.~(\ref{Full Hamiltonian simple form}) to describe light--matter interaction.
Also, we assume, that each molecule hosts one vibrational mode with the frequency $\omega_{\rm Vib}$.
For $j$-th molecule $\hat b_{{\rm Vib}j}^\dag$ (${\hat b_{{\rm Vib}j}}$) is the creation (annihilation) operator of the molecular vibration.
Here we approximate each molecular vibration by a harmonic oscillator.
The constant $\Lambda$ is the square root of the Huang--Rhys parameter~\cite{kirton2013nonequilibrium, cwik2016excitonic}.
Here we suppose that the electric field of the $\bf{k}$-th mode is distributed in the plane parallel to the mirrors according to $e^{i \bf{kr}}$.
Vector ${{\bf r}_j}$ points to the position of the $j$-th molecule, $\Omega_{j{\bf k}} =  - {{\bf{E}}_{{\bf k}}}{\bf d}_j/\hbar$ is a single-molecule Rabi frequency of the interaction with the cavity~\cite{scully1997quantum}, where ${\bf d}_j$ is the exciton transition dipole momentum of the molecule, ${\bf E}_{{\bf k}}$ is the electric field of the mode with in-plane momentum $\hbar {\bf k }$ inside the cavity.
We assume, that the dipole moments ${\bf d}_j$ are distributed randomly and evenly over the cavity.
Also, here we focus on the rather resonant condition $\omega_{\rm exc} - \omega_{{\rm Cav}|{\bf k}={\bf 0}} \sim \omega_{\rm Vib}$.
We list the main operators we use in the text in Table~\ref{table: operators}.

\begin{table}
\begin{center}
  \caption{The list of main operators}
\begin{tabular}{  c  c   } 
  \hline\hline
  Symbol & Meaning  \\ 
  \hline
  
  $\hat a_{{\rm Cav}|{\bf k}}$  & 
  Annihilation operator of a cavity photon \\
  & 
  with wave  vector $\bf k$ \\
  
  $\hat \sigma_{{\rm Exc}|j}$  & 
  Annihilation operator of an exciton at point ${\bf r}_j$ \\ 
  
  $\hat b_{{\rm Vib}|j}$  & 
  Annihilation operator of a molecular vibration \\  
    & 
  at point ${\bf r}_j$ \\ 
  
  $\hat s_{{\rm Low}|{\bf k}}$  & 
  Annihilation operator of a lower polariton with \\
  & 
  wave vector $\bf k$ \\
  
  $\hat s_{{\rm Up}|{\bf k}}$  & 
  Annihilation operator of an upper polariton \\ 
  & 
  with wave vector $\bf k$ \\
  
  $\hat c_{{\rm Vib}|{\bf q}}$  & 
  Annihilation operator of a bright molecular  \\ 
  & 
  vibration with wave vector $\bf q$ \\ 
    
  $\hat n_{\rm Exc_D}$  & 
  Number operator of dark excitons \\ 
     
  $\hat n_{\rm Vib_D}$  & 
  Number operator of dark molecular vibrations \\ 
      
  $\hat J_{\bf k}^{\rm (Bright)}$  & 
  Energy flow from bright excitons to lower \\ 
   & 
  polaritons with $\bf k$ around $\bf 0$ \\ 
       
  $\hat J_{\bf k}^{\rm (Dark)}$  & 
  Energy flow from dark excitons to lower \\ 
  & 
  polaritons with $\bf k$ around $\bf 0$ \\ 
        
  $\hat n_{{\rm Pol}|{\bf k}}$  & 
  $\hat s_{{\rm Low}|{\bf k}}^\dag \hat s_{{\rm Low}|{\bf k}}$ with $\bf k$ around $\bf 0$ \\ 
        
  $\hat n_{{\rm Exc}|{\bf k}_{ex}}$  & 
  Number operator of bright  \\ 
  & 
  excitons $\hat s_{{\rm Low}|{\bf k}_{ex}}^\dag \hat s_{{\rm Low}|{\bf k}_{ex}}$ \\ 

  $\hat n_{{\rm Vib}|{\bf q}}$  & 
  Number operator of bright vibrations $\hat c_{{\rm Vib}|{\bf q}}^\dag \hat c_{{\rm Vib}|{\bf q}}$ \\ 
  \hline\hline
\end{tabular}
  \label{table: operators}
\end{center}
\end{table}

In optical cavities, molecular electronic and vibrational states can hybridize with photons, forming new polaron-polariton eigenstates~\cite{kirton2013nonequilibrium, zasedatelev2021single, wu2016polarons, tereshchenkov2024thermalization}. Consequently, an accurate description of these systems requires consideration of both strong exciton-cavity and exciton-vibrational interactions.
We introduce the dressed excitonic and vibrational states with subsequent transition to the mixed light-matter states: upper and lower polaritons. Then we transform the Hamiltonian above Eq.~(\ref{Full Hamiltonian simple form}) into the basis of the light-matter states (see Appendix~\ref{appendix: Hamiltonian})
\begin{multline}\label{FullHamiltonian_polaritons}
\hat H = 
\sum\limits_{\bf k} 
\hbar \omega_{{\rm Up}|{\bf k}}
\hat s_{{\rm Up}|{\bf k}}^\dag \hat s_{{\rm Up}|{\bf k}}
+
\\
\sum\limits_{\bf k} 
\hbar \omega_{{\rm Low}|{\bf k}}
\hat s_{{\rm Low}|{\bf k}}^\dag \hat s_{{\rm Low}|{\bf k}}
+
\sum\limits_{\bf k} 
\hbar \omega_{\rm Vib}
\hat c_{{\rm Vib}|{\bf k}}^\dag \hat c_{{\rm Vib}|{\bf k}}
-
\\
\sum\limits_{{\bf k}, {\bf k}'} 
\frac{ \hbar \Lambda \Omega_R }{ \sqrt{N_{\rm mol}} }
\left[ 
\left(
\cos \varphi_{\bf k'}
\hat s_{{\rm Up}|{\bf k'}}^\dag
-
\sin \varphi_{\bf k'}
\hat s_{{\rm Low}|{\bf k'}}^\dag
\right)
\right. \\ \left.
\hat c_{{\rm Vib}|{\bf k'}-{\bf k }}
\left(
\cos \varphi_{\bf k} 
\hat s_{{\rm Low}|{\bf k}}
+
\sin \varphi_{\bf k} 
\hat s_{{\rm Up}|{\bf k}}
\right)
+
h.c.
\right]
+
\\
\sum\limits_{{\bf k}, {\bf k}'} 
\frac{ \hbar \Lambda \Omega_R}{\sqrt{N_{\rm mol}} }
\left[ 
\left(
\cos \varphi_{\bf k'}
\hat s_{{\rm Up}|{\bf k'}}^\dag
-
\sin \varphi_{\bf k'}
\hat s_{{\rm Low}|{\bf k'}}^\dag
\right)
\right. \\ \left.
\hat c_{{\rm Vib}|{\bf k}-{\bf k' }}^\dag
\left(
\cos \varphi_{\bf k} 
\hat s_{{\rm Low}|{\bf k}}
+
\sin \varphi_{\bf k} 
\hat s_{{\rm Up}|{\bf k}}
\right)
+
h.c.
\right]
+
\\
N_{\rm mol} \hbar \omega_{\rm Exc} \hat n_{\rm Exc_D}
+
N_{\rm mol} \hbar \omega_{\rm Vib} \hat n_{\rm Vib_D}
+
\hat H_{\rm cav-dark},
\end{multline}
where $\Omega_R$ is the collective Rabi frequency, 
\begin{multline} \label{TransformationAngle}
\varphi_{\bf k}=
{\rm arctg}
\Bigg[ 
\sqrt{
\frac
{(\omega_{\rm Exc}-\omega_{{\rm Cav}|{\bf k}})^2+4\Omega_R^2}
{4\Omega_R^2}
}
 \\ 
-
\frac{\omega_{\rm Exc}-\omega_{{\rm Cav}|{\bf k}}}{2\Omega_R}
\Bigg]
\end{multline}
is parameter determining the Hopfield coefficients $\sin\varphi_{\bf k}$ and $\cos\varphi_{\bf k}$, $\omega_{\rm Exc} = \omega_{\rm exc} - \Lambda^2 \omega_{\rm Vib}$ is the energy of the dressed exciton states.
The transition from Hamiltonian~(\ref{Full Hamiltonian simple form}) to Hamiltonian~(\ref{FullHamiltonian_polaritons}) involves the introduction of the collective bright states, namely upper and lower polaritons as well as bright molecular vibrations defined by the following operators
\begin{multline} 
\hat s_{{\rm Low}|{\bf k}} 
= 
\hat a_{{\rm Cav}|{\bf k}} \cos \varphi_{{\bf k}} 
- 
\\
\frac{1}{\Omega_R}
\sum_{j=1}^{N_{\rm mol}}  
\Omega_{j{\bf k}}
\hat \sigma_{{\rm Exc}j} e^{\Lambda (\hat b_{{\rm vib}j}^\dag + \hat b_{{\rm vib}j})} 
{e^{-i{\bf k}{\bf r}_j}} \sin \varphi_{{\bf k}},
\end{multline}
\begin{multline} 
\hat s_{{\rm Up}|{\bf k}} 
= 
\hat a_{{\rm Cav}|{\bf k}} \sin \varphi_{{\bf k}} 
+
\\
\frac{1}{\Omega_R}
\sum_{j=1}^{N_{\rm mol}}  
\Omega_{j{\bf k}}
\hat \sigma_{{\rm Exc}j} e^{\Lambda (\hat b_{{\rm Vib}j}^\dag + \hat b_{{\rm Vib}j})} 
{e^{-i{\bf k}{\bf r}_j}} \cos \varphi_{{\bf k}},
\end{multline}
\begin{equation} 
\hat c_{{\rm Vib}|{\bf k}} = 
\frac{1} {\sqrt{N_{\rm mol}}}
\sum_{j=1}^{N_{\rm mol}} 
\left(
\hat b_{{\rm Vib}j} 
+
\Lambda \hat \sigma^\dag_{{\rm Exc}j} \hat \sigma_{{\rm Exc}j} 
\right)
e^{-i{\bf k}{\bf r}_j}.
\end{equation}

\begin{figure}
\includegraphics[width=0.95\linewidth]{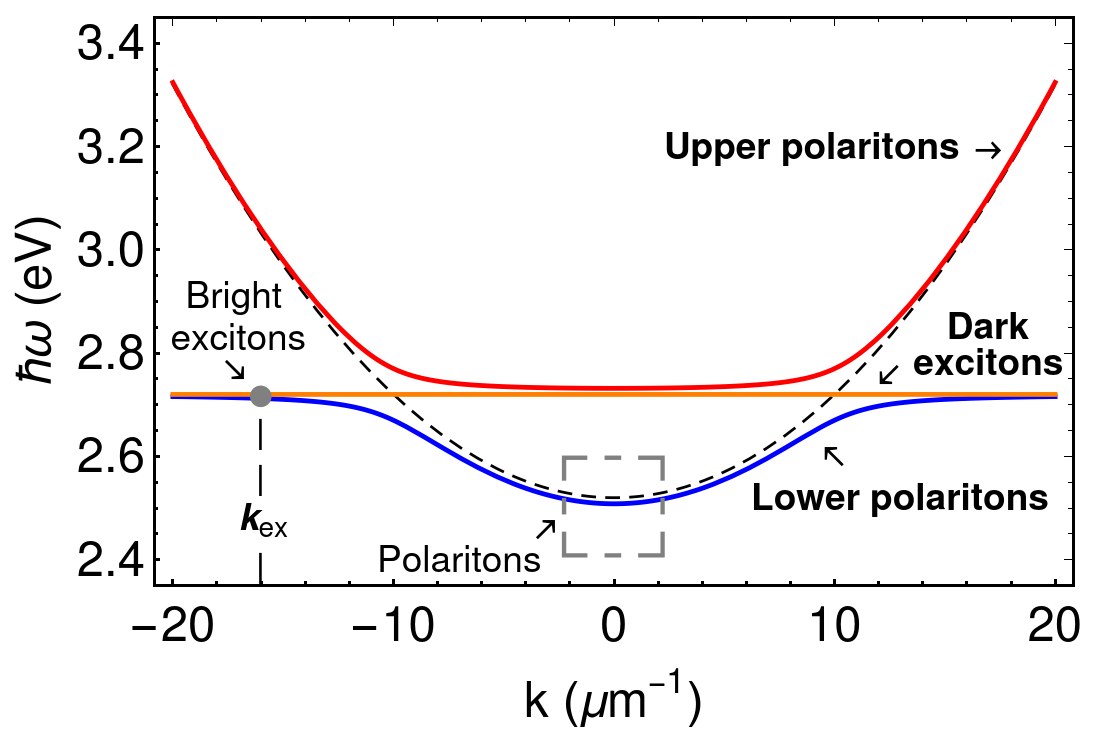}
\caption{
Dispersion relations of the upper polaritons (red line), lower polaritons (blue line) and dark exciton states (orange line). The bare cavity mode is illustrated by the black dashed line.
Here we use the following parameters of the system: $\Omega_R = 0.05$~$\rm eV$,  $\omega_{{\rm Cav}|{\bf k}} = \omega_{{\rm Cav}|{\bf k}={\bf 0}} + \alpha_{\rm Cav} {\bf k}^2$ with $\omega_{{\rm Cav}|{\bf k}={\bf 0}} = 2.52~{\rm eV}$, $\alpha_{\rm Cav} = 2 \cdot 10^{-3}$~${\rm eV} \mu {\rm m^2}$ and $\omega_{\rm Exc} = 2.72~{\rm eV}$, in agreement with the recent experiments~\cite{zasedatelev2019room, zasedatelev2021single}.
}
    \label{fig:dispersion}
\end{figure}

Upper and lower polaritons and bright molecular vibrations are phase-coherent, many-body delocalized states with a well-defined in-plane momentum $\hbar\bf k$, matching the corresponding eigenstate of the cavity. 
The dispersions of the lower and upper polaritons are
\begin{equation} \label{FrequenciesOfLowerPolaritons}
\omega _{{\rm Low}|{\bf k}} 
= 
\frac{\omega_{\rm Exc} + \omega_{{\rm Cav}|{\bf k}} }{ 2}
-
\sqrt{
\frac{
\left( 
\omega_{\rm Exc} - \omega _{{\rm Cav}|{\bf k}} 
\right)^2 
}{
4
}
+
\Omega_R^2
},
\end{equation}
\begin{equation} \label{FrequenciesOfUpperPolaritons}
\omega _{{\rm Up}|{\bf k}} 
= 
\frac{\omega_{\rm Exc} + \omega_{{\rm Cav}|{\bf k}} }{ 2}
+
\sqrt{
\frac{
\left( 
\omega_{\rm Exc} - \omega _{{\rm Cav}|{\bf k}} 
\right)^2 
}{ 4
}
+
\Omega_R^2
}.
\end{equation}
Here we assume that lower and upper polaritons are bosonic particles.
This assumption was justified for large number of molecules and low average occupations of the exciton states per molecule in Ref.\cite{combescot2008microscopic}.

Fig.~\ref{fig:dispersion} shows dispersion relations for the polariton and dark exciton states. 
It is convenient to divide the lower polaritons into bright excitons and polaritons based on wave vector as shown on Fig.~\ref{fig:dispersion}.
Besides the phase-coherent bright states there is also a manifold of localized states, so-called dark excitons and dark vibrations, those are lacking well-defined momentum, see Appendix A for details.
The operators for the number of dark excitons $\hat n_{\rm Exc_D}$, and dark vibrations $\hat n_{\rm Vib_D}$ per molecule are defined by Eq.~(\ref{H_dark_exc})~and Eq.~(\ref{H_dark_vib}).

\begin{figure}
\includegraphics[width=1\linewidth]{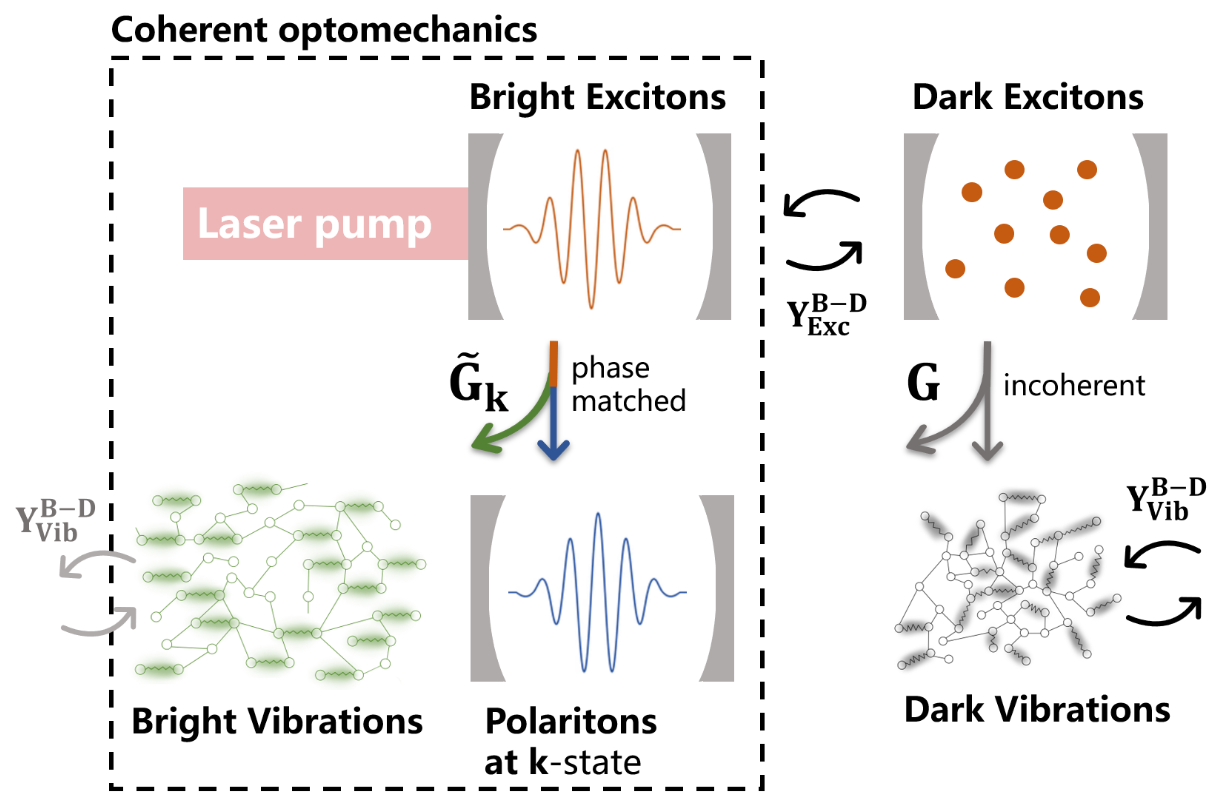}
\caption{
Schematic of the main subsystems, namely collective phase-coherent excitonic and vibrational states, polaritons; as well as localized dark excitons and molecular vibrations. The diagram illustrates the main energy exchange mechanisms including the coherent $\tilde{G}_{\bf k}$ and incoherent $G$ optomechanical interactions, the interaction between the bright and dark excitons  $\gamma^{\rm B-D}_{\rm Exc}$, as well as the interaction between the bright and dark vibrations $\gamma^{\rm B-D}_{\rm Vib}$.
}
    \label{fig:Layout}
\end{figure}

The tripartite interaction between the bright states in Hamiltonian~(\ref{FullHamiltonian_polaritons}) results in the coherent energy exchange between the bright excitons, polaritons, and bright vibrations as shown schematically in Fig.~\ref{fig:Layout}.
The process is nonlinearly dependent on the occupation of the bright states involved and can be characterized by constant $G_{\bf k}$~(see Fig.~\ref{fig:Layout}), which we rigorously introduce below in Eq.~(\ref{optomechanical_constant_2}).
This constant is the direct analog of the optomechanical damping rate~\cite{aspelmeyer2014cavity} and is proportional to both the square of the Huang--Rhys factor, $\Lambda^2$, and the square of the Rabi frequency, $\Omega_R^2$.

Although dark excitons do not directly interact with the cavity to form polariton states, they significantly influence polariton dynamics~\cite{herrera2017dark}. The interaction between lower polaritons, dark excitons, and dark molecular vibrations is described by Hamiltonian $\hat H_{\rm cav-dark}$ within Eq.~(\ref{FullHamiltonian_polaritons}). 
This interaction is resonant, involving the annihilation of a dark exciton, the creation of a lower polariton, and the emission of a dark vibration, as depicted in Fig.~\ref{fig:Layout}. The resultant energy transfer also shows a nonlinear dependence on the occupation of these states and is quantified by the constant $G$ introduced in Eq.~(\ref{optomechanical_constant}). Given that the total number of dark exciton states substantially exceeds that of the bright states~\cite{tereshchenkov2024thermalization}, this process becomes increasingly significant below the condensation threshold.

The system under consideration is inherently an open quantum system.
It dissipates the energy to the environment and gets the energy from the external laser drives. 
We use Born--Markov approximation to describe all the relaxation processes including polariton thermalization. 
Here, we mostly focus on the blue-detuned resonant pumping conditions~\cite{zasedatelev2021single, shishkov2023mapping} where the laser excites bright excitonic states with the rate $\varkappa_{\rm Pump}$ at high wavevector ${\bf k}={\bf k}_{\rm ex}$ as shown in Fig.~\ref{fig:dispersion}.
The dissipation rate of dark excitons is $\gamma_{\rm Exc}$ and the corresponding dephasing rate is $\Gamma_{\rm Exc}$.
Both dark and bright molecular vibrations have the same dissipation rate $\gamma_{\rm Vib}$.
The dissipation rates of the lower polariton states are $\gamma_{{\rm Low}|{\bf k}}$ and for upper polariton states are $\gamma_{{\rm Up}|{\bf k}}$.
The formal introduction of dissipation processes through Lindblad superoperators can be found in Appendix~\ref{appendix: Relaxation}.
The typical quality factor (Q) of high-energy vibrational modes in molecular systems is approximately $Q \sim 100$, as observed in both spontaneous Raman and IR spectra~\cite{coles2011vibrationally,zasedatelev2019room, ahn2018vibrational, simpkins2023control}. This implies vibrational dephasing rates of approximately $1~{\rm meV}$. We use $\gamma_{\rm Vib} = 3~{\rm meV}$, which is representative of C-C stretching modes in conjugated molecular systems (for example, polymer MeLPPP, see Fig. 2 in SM to Ref.~\cite{zasedatelev2021single}). Note, vibrational/phonon modes in TMDC exhibit similar Q-factors of $\sim100$ as evidenced in their Raman spectra~\cite{dewambrechies2023enhanced}.

Recent microscopic theories have indicated that polariton thermalization primarily arises from low-energy molecular vibrations that are coupled to the material component of polaritons~\cite{tereshchenkov2024thermalization}. The thermalization rates for upward and downward transitions among the lower polariton branch, denoted as $\gamma_{\rm therm}^{{{\bf k}_2}{{\bf k}_1}}$, for states with wavevectors ${\bf k}_1$ and ${\bf k}_2$, respectively, are connected through the Kubo--Martin--Schwinger relation, as detailed in Appendix B Eq.~(\ref{kubo-martin-schwinger relation}). Since the total number of dark exciton states greatly exceeds the number of bright states, dark excitons effectively act as a reservoir for the bright excitons. We account for this mechanism phenomenologically, assuming that the corresponding relaxation rate is $\gamma_{\rm Exc}^{\rm B-D} = \Gamma_{\rm Exc}$ as shown in Fig.~\ref{fig:Layout}.
A similar principle is applied to the bright and dark vibrational states where the corresponding relaxation rate $\gamma_{\rm Vib}^{\rm B-D} = \gamma_{\rm Vib}/2$.

\section{Coherent optomechanics}\label{sec:susceptibility}

In the following Section we focus on the coherent part of the system. Here we analyse the Hamiltonian and develop optomechanical formalism for the coherent light-matter and vibrational states. 

We assume the laser drive initially populates bright exciton state with wave vector ${\bf k}_{\rm ex}$. On the later stages, the population partly reaches polariton states via coherent interaction with vibrations and through the incoherent dark states as shown in Fig.~\ref{fig:Layout}. Note, this is a single-beam technique in contrast to coherent Raman methods discussed in Section~\ref{sec:consequences}. Eventually polaritons can undergo nonequilibrium Bose--Einstein condensation when the laser drive is above critical threshold~\cite{shishkov2023mapping}. The BEC manifests itself in macroscopic occupation of a single polariton state with in-plane momentum $\hbar\bf{k}={\bf 0}$ accompanied by the build-up of the coherence and long-range order~\cite{deng2010exciton}. In the following we restrict our analysis to the regime above condensation threshold.  
We assume that the system is pumped in the CW regime, i.e. the duration of the pumping surpasses all the relaxation times in the system listed in Table~\ref{table: parameters}. While this assumption might seem hard to attain experimentally at first glance, the fast dynamics of polariton systems allow reaching the steady state within a few tens to hundreds of picoseconds after the excitation. This makes it feasible to drive the system into a quasi-steady-state regime with nanosecond excitation. Polariton condensation driven by a quasi-CW excitation has been realized experimentally~\cite{putintsev2020nano}.

The method above can be further extended by the stimulated polariton condensation with a two-beam approach~\cite{zasedatelev2019room,zasedatelev2021single} providing flexibility upon the choice of the energy and momentum for both the polariton and vibrational state invloved~\cite{baranikov2020all}. 



In this optomechanical picture, we do not distinguish between bright excitons and polaritons, instead addressing the entire lower polariton dispersion branch (see Fig.~\ref{fig:dispersion}).
We consider coherent part of Hamiltonian~(\ref{FullHamiltonian_polaritons}) excluding the  dark states contributing to the incoherent dynamics.   
\begin{multline} \label{H_bright}
\hat H_{\rm bright} = 
\sum\limits_{\bf k} 
\hbar \omega_{{\rm Low}|{\bf k}}
\hat s_{{\rm Low}|{\bf k}}^\dag \hat s_{{\rm Low}|{\bf k}}
+
\\
\sum\limits_{\bf k} 
\hbar \omega_{\rm Vib}
\hat C_{{\rm Vib}|{\bf k}}^\dag \hat C_{{\rm Vib}|{\bf k}}
+
\\
\sum\limits_{{\bf k}, {\bf k}'} 
\hbar g_{\bf kk'} 
\hat s_{{\rm Low}|{\bf k'}}^\dag
\hat s_{{\rm Low}|{\bf k}}
\left(
\hat C_{{\rm Vib}|{\bf k'}-{\bf k }}
+
\hat C_{{\rm Vib}|{\bf k}-{\bf k '}}^\dag
\right),
\end{multline}
where we denote $\hat C_{{\rm Vib}|{\bf k}} = -i \hat c_{{\rm Vib}|{\bf k}}$ and
\begin{equation}
g_{\bf kk'} = i \frac{ \Lambda \Omega_R }{ \sqrt{N_{\rm mol}} }
\sin(\varphi_{\bf k'}-\varphi_{\bf k}).
\end{equation}

The Hamiltonian Eq.~(\ref{H_bright}) governs the Hermitian dynamics of the lower polaritons and bright molecular vibrations. This Hamiltonian is analogous to those found in well-established multimode cavity optomechanics~\cite{aspelmeyer2014cavity}, though the modes involved differ significantly in nature. In our system, the ``optical'' modes emerge from the strong light-matter interactions and consist of matter components, while the ``mechanical'' modes, represented by collective molecular vibrations, which are delocalized across an ensemble of molecules. These mechanical modes are tightly bounded to the polariton states through strong exciton-vibration interactions, resulting in the polariton optomechanical coupling constant $g_{\bf kk'}$.

\begin{table*}
\begin{center}
\caption{ The typical parameters used in the calculations.}
\begin{tabular}{  c  c  c  c  } 
  \hline\hline
  Symbol & Meaning & Value & Reference \\

  $\omega_{{\rm Cav}|{\bf k}}$  & 
  \makecell{ Energy of a cavity photon \\ with wave vector $\bf k$} & 
  $2.52~{\rm eV}~{\rm at}~{{\bf k}={\bf 0}}$ & 
  \cite{baranikov2020all, zasedatelev2021single} \\ 

  $\alpha_{\rm Cav}$  & 
  \makecell{Dispersion coefficient \\ of the cavity photons} & 
  $2~{\rm meV \cdot \mu m^2}$ & 
  \cite{baranikov2020all, zasedatelev2021single} \\ 

  $\omega_{\rm Exc}$  & 
  Energy of a dressed exciton & 
  $2.72~{\rm eV}$ & 
  \cite{plumhof2014room, zasedatelev2019room, zasedatelev2021single, xia2023ladder} \\
  
  $\Omega_R$  & 
  \makecell{Rabi energy of cavity \\ photons and excitons} & 
  $50~{\rm meV}$ & 
  \cite{plumhof2014room, sanvitto2016road, zasedatelev2019room, zasedatelev2021single} \\

  $\omega_{\rm Vib}$  & 
  Energy of a vibrational quanta & 
  $200~{\rm meV}$ & 
  \cite{zasedatelev2019room, zasedatelev2021single, guha2003temperature, coropceanu2002hole, xia2023ladder} \\

  $\Lambda^2$  & 
  Huang-Rhys factor & 
  $1$ & 
  \cite{guha2003temperature, tereshchenkov2024thermalization, kena2008strong, kena2010room, coropceanu2002hole} \\ 

  $N_{\rm mol}$  & 
  Total number of the molecules & 
  $10^8$ & 
  \cite{zasedatelev2021single} \\ 

  $\omega_{{\rm Low}|{\bf k}} = \omega_{{\rm Pol}|{\bf k}}$  & 
  \makecell{Energy of a lower polariton \\ with wave vector $\bf k$} & 
  $2.51~{\rm eV}~{\rm at}~{{\bf k}={\bf 0}}$ & 
  \cite{baranikov2020all, zasedatelev2021single} \\ 
    
  $\alpha_{\rm pol}$  & 
  \makecell{Dispersion of lower polaritons \\ around ${\bf k} = {\bf 0}$} & 
  $1.88~{\rm meV \cdot \mu m^2}$ & 
  \cite{baranikov2020all, zasedatelev2021single} \\ 

  $\omega_{{\rm Up}|{\bf k}}$  & 
  \makecell{Energy of an upper polariton \\ with wave vector $\bf k$} & 
  $2.73~{\rm eV}~{\rm at}~{{\bf k}={\bf 0}}$ & 
  \cite{zasedatelev2019room, zasedatelev2021single} \\ 
  
  $\sin^2\varphi_{\bf k}$, $\cos^2\varphi_{\bf k}$  & 
  \makecell{Hopfield coefficients for polaritons \\ with wave vector $\bf k$} & 
  $\in (0,1)$ & 
  \cite{plumhof2014room, zasedatelev2021single} \\ 

  $k_{\rm ex}$  & 
  Wave vector of an external field & 
  Varies in Sec.~\ref{sec:susceptibility},  $16~{\rm \mu m^{-1}}$ in Sec.~\ref{sec:consequences} & 
  \cite{zasedatelev2019room, zasedatelev2021single} \\ 

  $G_{\bf k}$  & 
  \makecell{Effective optomechanical \\ constant for dark states} & 
  \makecell{$0.5~{\rm eV}$ at ${\bf k}={\bf 0}$ \\ and $k_{\rm ex}=16~{\rm \mu m^{-1}}$} & 
  Eq.~(\ref{optomechanical_constant}) \\ 

  $\tilde G_{{\bf k}_{\rm ex}{\bf k}}$  & 
  \makecell{Effective optomechanical \\ constant for bright states} & 
  \makecell{$0.5~{\rm eV}$ at ${\bf k}={\bf 0}$ \\ and $k_{\rm ex}=16~{\rm \mu m^{-1}}$} & 
   Eq.~(\ref{optomechanical_constant_2}) \\

  $\varkappa_{\rm Pump}$  & 
  Pumping parameter & 
  Varies & 
   \\   
   
  $\gamma_{{\rm Cav}|{\bf k}}$  & 
  \makecell{Dissipation rate of cavity \\ photons with wave vector $\bf k$} & 
  $2.5~{\rm meV}$ & 
  \cite{zasedatelev2019room, zasedatelev2021single, betzold2019coherence} \\ 
        
  $\gamma_{{\rm Exc}}$  & 
  Dissipation rate of excitons & 
  $0.01~{\rm meV}$ & 
  \cite{plumhof2014room, zasedatelev2019room, zasedatelev2021single, xia2023ladder} \\ 
        
  $\Gamma_{{\rm Exc}}$  & 
  Dephasing rate of excitons & 
  $10~{\rm meV}$ & 
  \cite{kena2008strong, zasedatelev2019room, zasedatelev2021single} \\ 
        
  $\gamma_{{\rm Vib}}$  & 
  \makecell{Dissipation rate of \\ molecular vibrations} & 
  $2~{\rm meV}$ & 
  \cite{zasedatelev2019room, zasedatelev2021single, xia2023ladder} \\ 
            
  $\gamma_{{\rm Low}|{\bf k}}=\gamma_{{\rm Pol}|{\bf k}}$  & 
  \makecell{Dissipation rate of lower polaritons \\ with wave vector $\bf k$} & 
  $2.5~{\rm meV}$ at ${\bf k}={\bf 0}$ & 
  \cite{baranikov2020all, zasedatelev2021single, betzold2019coherence} \\ 
     
  $\gamma_{\rm Exc}^{\rm B-D}$  & 
  \makecell{Trastition rate from \\ bright to dark excitons} & 
  $\Gamma_{\rm Exc}$ & 
  \cite{perez2023simulating} \\ 
      
  $\gamma_{\rm Vib}^{\rm B-D}$  & 
  \makecell{Trastition rate from bright \\ to dark molecular vibrations} & 
  $\gamma_{\rm Vib}/2$ & 
   \\ 
      
  $\gamma_{\rm therm}^{\bf kk'}$  & 
  \makecell{Thermalization rate between lower \\ polaritons with wave vectors $\bf k$ and $\bf k'$} & 
  $0.01~{\rm meV}$ & 
  \cite{tereshchenkov2024thermalization} \\ 
  \hline\hline
        
\end{tabular}
  \label{table: parameters}
\end{center}
\end{table*}

Like the standard optomechanical Hamiltonian that describes the interaction of a movable mirror with the cavity radiation field is fundamentally nonlinear involving three operators, the genuine polariton optomechanical interaction is also tripartite. However, given our primary focus is on optomechanics with light-matter BEC well above condensation threshold the interaction term in Eq.~(\ref{H_bright}) can be effectively linearized around the average polariton amplitude  $\langle\hat s_{{\rm Low}{\bf k}_{\rm ex}}\rangle = \sqrt{n_{{\rm Low}{\bf k}_{\rm ex}}}e^{-i\omega_{{\rm Low}|{\bf k}_{\rm ex}}t}$.
Introduction of new operators $\hat s_{{\rm Low}|{\bf k}} = \hat S_{{\rm Low}|{\bf k}}e^{-i\omega_{{\rm Low}|{\bf k}_{\rm ex}}t}$ leads to 

\begin{multline} \label{H_bright_eff}
\hat H_{\rm bright}^{\rm lin} = 
-\sum\limits_{\bf k} 
\hbar \Delta \omega_{{\bf k}_{\rm ex}{\bf k}}
\hat S_{{\rm Low}|{\bf k}}^\dag 
\hat S_{{\rm Low}|{\bf k}}
+
\\
\sum\limits_{\bf k} 
\hbar \omega_{\rm Vib}
\hat C_{{\rm Vib}|{\bf k}}^\dag \hat C_{{\rm Vib}|{\bf k}}
+
\\
\sum\limits_{{\bf k}} 
\hbar g_{{\bf k}{\bf k}_{\rm ex}} 
\sqrt{n_{{\rm Low}{\bf k}_{\rm ex}}}
\hat S_{{\rm Low}|{\bf k}}
\left(
\hat C_{{\rm Vib}|{\bf k}_{\rm ex}-{\bf k }}
+
\hat C_{{\rm Vib}|{\bf k}-{\bf k }_{\rm ex}}^\dag
\right)
+
\\
\sum\limits_{{\bf k}} 
\hbar g_{{\bf k}_{\rm ex}{\bf k}} 
\sqrt{n_{{\rm Low}{\bf k}_{\rm ex}}}
\hat S_{{\rm Low}|{\bf k}}^\dag
\left(
\hat C_{{\rm Vib}|{\bf k}-{\bf k }_{\rm ex}}
+
\hat C_{{\rm Vib}|{\bf k}_{\rm ex}-{\bf k}}^\dag
\right),
\end{multline}
where $\Delta \omega_{{\bf k}_{\rm ex}{\bf k}} = \omega_{{\rm Low}|{\bf k}_{\rm ex}} - \omega_{{\rm Low}|{\bf k}}$.
The optomechanical coupling strength depends on the wave vectors ${\bf k}_{\rm ex}$ and $\bf k$.
In the case, $\Delta \omega_{{\bf k}_{\rm ex}{\bf k}} \approx \omega_{\rm Vib}$, the vacuum optomechanical strength is $g_{{\bf k}_{\rm ex}{\bf k}}$ and net one is $g_{{\bf k}_{\rm ex}{\bf k}} \sqrt{n_{{\rm Low}|{{\bf k}_{\rm ex}}}}$. 

We proceed further with the linear analysis of the coupled polariton and vibrational modes. To determine optomechanical response of the system, we consider a harmonic weak test force $\hat f_{\bf k}$ acting on the bright molecular vibrations~\cite{aspelmeyer2014cavity}.
This interaction is set by the Hamiltonian of the weak test force
\begin{equation}
\hat H_{\rm test} = 
\sum_{\bf k} 
\left(
\hat f_{\bf k}e^{-i\omega t} \hat C_{\bf k}^\dag
+
\hat f_{\bf k}^\dag e^{i\omega t} \hat C_{\bf k}
\right).
\end{equation}

In the next step, we derive the equations of motion for complex amplitudes of polariton and vibrational states under the linearized optomechanical interaction. 
The optomechanical Hamiltonian Eq.~(\ref{H_bright_eff}) describes the process where a polariton with wave vector ${\bf k}_{\rm ex}$ is annihilated and a polariton with wave vector $\bf k$ is created, accompanied by the creation of the bright molecular vibration with wave vector ${\bf k}_{\rm ex}-{\bf k}$.
Here, we are mostly interested in the properties of this bright molecular vibration. 
Therefore, we set the weak test force $\hat f_{\bf k}$ acting only on the molecular vibrations with the wave vector ${\bf k}_{\rm ex}-{\bf k}$.
As the result we obtain the equations of motion for the complex amplitudes

\begin{multline} \label{eq of motion 1}
\frac{ d\hat S_{{\rm Low}|{\bf k}} }{dt} = 
\left(
i\Delta \omega_{{\bf k}_{\rm ex}{\bf k}}-\frac{\Gamma_{{\rm Low}|{\bf k}}}{2}
\right)
\hat S_{{\rm Low}|{\bf k}}
-
\\
i g_{{\bf k}_{\rm ex}{\bf k}} 
\sqrt{n_{{\rm Low}{\bf k}_{\rm ex}}}
\left(
\hat C_{{\rm Vib}|{\bf k}-{\bf k }_{\rm ex}}
+
\hat C_{{\rm Vib}|{\bf k}_{\rm ex}-{\bf k}}^\dag
\right),
\end{multline}
\begin{multline} \label{eq of motion 2}
\frac{ d\hat C_{{\rm Vib}|{\bf k}-{\bf k }_{\rm ex}} }{dt} =
\left(
-i\omega_{\rm Vib}-\frac{\gamma_{\rm Vib}}{2}
\right) 
\hat C_{{\rm Vib}|{\bf k}-{\bf k }_{\rm ex}} - \\
i g_{{\bf k}{\bf k}_{\rm ex}} 
\sqrt{n_{{\rm Low}{\bf k}_{\rm ex}}}
\hat S_{{\rm Low}|{\bf k}} -
i g_{{\bf k}_{\rm ex}{\bf Q}} 
\sqrt{n_{{\rm Low}{\bf k}_{\rm ex}}}
\hat S_{{\rm Low}|{\bf Q}}^\dag,
\end{multline}
\begin{multline} \label{eq of motion 3}
\frac{ d\hat S_{{\rm Low}|{\bf Q}}^\dag }{dt} =
\left(
-i\Delta \omega_{{\bf k}_{\rm ex}{\bf Q}}-\frac{\Gamma_{{\rm Low}|{\bf Q}}}{2}
\right) 
\hat S_{{\rm Low}|{\bf Q}}
+
\\
i g_{{\bf Q}{\bf k}_{\rm ex}} 
\sqrt{n_{{\rm Low}{\bf k}_{\rm ex}}}
\left(
\hat C_{{\rm Vib}|{\bf k }-{\bf k}_{\rm ex}}
+
\hat C_{{\rm Vib}|{\bf k }_{\rm ex}-{\bf k}}^\dag
\right),
\end{multline}
\begin{multline} \label{eq of motion 4}
\frac{ d\hat C_{{\rm Vib}|{\bf k }_{\rm ex}-{\bf k}}^\dag }{dt} =
\left(
i\omega_{\rm Vib}-\frac{\gamma_{\rm Vib}}{2}
\right) 
\hat C_{{\rm Vib}|{\bf k }_{\rm ex}-{\bf k}}^\dag  + \\
i g_{{\bf k}{\bf k}_{\rm ex}} 
\sqrt{n_{{\rm Low}{\bf k}_{\rm ex}}}
\hat S_{{\rm Low}|{\bf k}} +
i g_{{\bf k}_{\rm ex}{\bf Q}} 
\sqrt{n_{{\rm Low}{\bf k}_{\rm ex}}}
\hat S_{{\rm Low}|{\bf Q}}^\dag +
\\
i \hat f_{{\bf k }_{\rm ex}-{\bf k}}^\dag e^{i\omega t},
\end{multline}
where we denote the wave vector ${\bf Q}=2{\bf k}_{\rm ex}-{\bf k}$ and $\Gamma_{{\rm Low}|{\bf k}} = \Gamma_{\rm exc}\sin^2\varphi_{\bf k} + \gamma_{{\rm Low}|{\bf k}}$.

Here we neglect all fluctuations arising from thermal and quantum noise, instead we focus on the coherent linear response of the system to an external force. From Eq.~(\ref{eq of motion 1})--(\ref{eq of motion 4}) one can see immediately that the mechanical motion changes the damping and shifts the frequency of the polariton state, which in turn results in a change of polariton density at the BEC, thus acting back on the mechanical motion of the collective molecular vibrations. This kind of feedback loop is known as optomechanical backaction~\cite{aspelmeyer2014cavity}. 
To understand how optomechanical backaction alters the response of the vibrational subsystem, we transition to the frequency space and express the modified mechanical susceptibility, $\chi_{\rm Vib,eff}(\omega)$, defined by $\hat C_{{\bf k}_{\rm ex}-{\bf k}} = \chi_{\rm Vib,eff}(\omega) \hat f_{{\bf k}_{\rm ex}-{\bf k}}e^{-i\omega t}$~\cite{renninger2018bulk}
\begin{equation}
\chi_{\rm Vib,eff}^{-1}(\omega)
=
\chi_{\rm Vib}^{-1}(\omega)
+
\Sigma_{\rm Vib}(\omega),
\end{equation}
where $\chi_{\rm Vib}(\omega) = (\omega - \omega_{\rm Vib} + i\gamma_{\rm Vib}/2)^{-1}$ is the susceptibility of molecular vibrations in the abscence of optomechanical coupling. The optomechanical contribution to the linear response to an external force is expressed as follows:
\begin{multline}
\Sigma_{\rm Vib}(\omega) = 
\frac
{
|g_{{\bf k}_{\rm ex}{\bf k}}|^2 n_{{\rm Low}|{{\bf k}_{\rm ex}}}
}{
\omega - \Delta \omega_{{\bf k}_{\rm ex}{\bf k}} +
i\Gamma_{{\rm Low}|{\bf k}}/2
}
-
\\
\frac
{
|g_{{\bf k}_{\rm ex}{\bf Q}}|^2 n_{{\rm Low}|{{\bf k}_{\rm ex}}}
}{
\omega + \Delta \omega_{{\bf k}_{\rm ex}{\bf Q}} + 
i\Gamma_{{\rm Low}|{\bf Q}}/2
} .
\end{multline}
Being dependent on both $\bf k$ and ${\bf k}_{\rm ex}$, the susceptibility $\chi_{\rm Vib,eff}(\omega)$ allows us to determine the optomechanical frequency shift, $\Delta \omega_{{\rm Vib}|{\bf k}_{\rm ex},{\bf k}} = -{\rm Re}\Sigma_{\rm Vib}(\omega_{\rm Vib})$, and optomechanical damping $\Delta \gamma_{{\rm Vib}|{\bf k}_{\rm ex},{\bf k}} = 2{\rm Im}\Sigma_{\rm Vib}(\omega_{\rm Vib})$ of the bright molecular vibrations with wave vector ${\bf k}_{\rm ex}-{\bf k}$
\begin{multline}  \label{optomechanical_shift}
\Delta \omega_{{\rm Vib}|{\bf k}_{\rm ex},{\bf k}}
=
-
\frac
{
|g_{{\bf k}_{\rm ex}{\bf k}}|^2 n_{{\rm Low}|{{\bf k}_{\rm ex}}}
\left( \omega_{\rm Vib} - \Delta \omega_{{\bf k}_{\rm ex}{\bf k}} \right)
}{
\left( \omega_{\rm Vib} - \Delta \omega_{{\bf k}_{\rm ex}{\bf k}} \right)^2 + 
\left( \Gamma_{{\rm Low}|{\bf k}}/2\right)^2
} +
\\
\frac
{
|g_{{\bf k}_{\rm ex}{\bf Q}}|^2 n_{{\rm Low}|{{\bf k}_{\rm ex}}}
\left( \omega_{\rm Vib} + \Delta \omega_{{\bf k}_{\rm ex}{\bf Q}}   \right)
}{
\left( \omega_{\rm Vib} + \Delta \omega_{{\bf k}_{\rm ex}{\bf Q}} \right)^2 + 
\left( \Gamma_{{\rm Low}|{\bf Q}}/2\right)^2
} 
,
\end{multline}
\begin{multline} \label{optomechanical_damping}
\Delta \gamma_{{\rm Vib}|{{\bf k}_{\rm ex},{\bf k}}}
=
-
\frac
{
|g_{{\bf k}_{\rm ex}{\bf k}}|^2 n_{{\rm Low}|{{\bf k}_{\rm ex}}}
\Gamma_{{\rm Low}|{\bf k}}
}{
\left( \omega_{\rm Vib} - \Delta \omega_{{\bf k}_{\rm ex}{\bf k}} \right)^2 + 
\left( \Gamma_{{\rm Low}|{\bf k}}/2\right)^2
} 
+
\\
\frac
{
|g_{{\bf k}_{\rm ex}{\bf Q}}|^2 n_{{\rm Low}|{{\bf k}_{\rm ex}}}
\Gamma_{{\rm Low}|{\bf Q}}
}{
\left( \omega_{\rm Vib} + \Delta \omega_{{\bf k}_{\rm ex}{\bf Q}} \right)^2 + 
\left( \Gamma_{{\rm Low}|{\bf Q}}/2\right)^2
} 
.
\end{multline}

\begin{figure}
\includegraphics[width=0.95\linewidth]{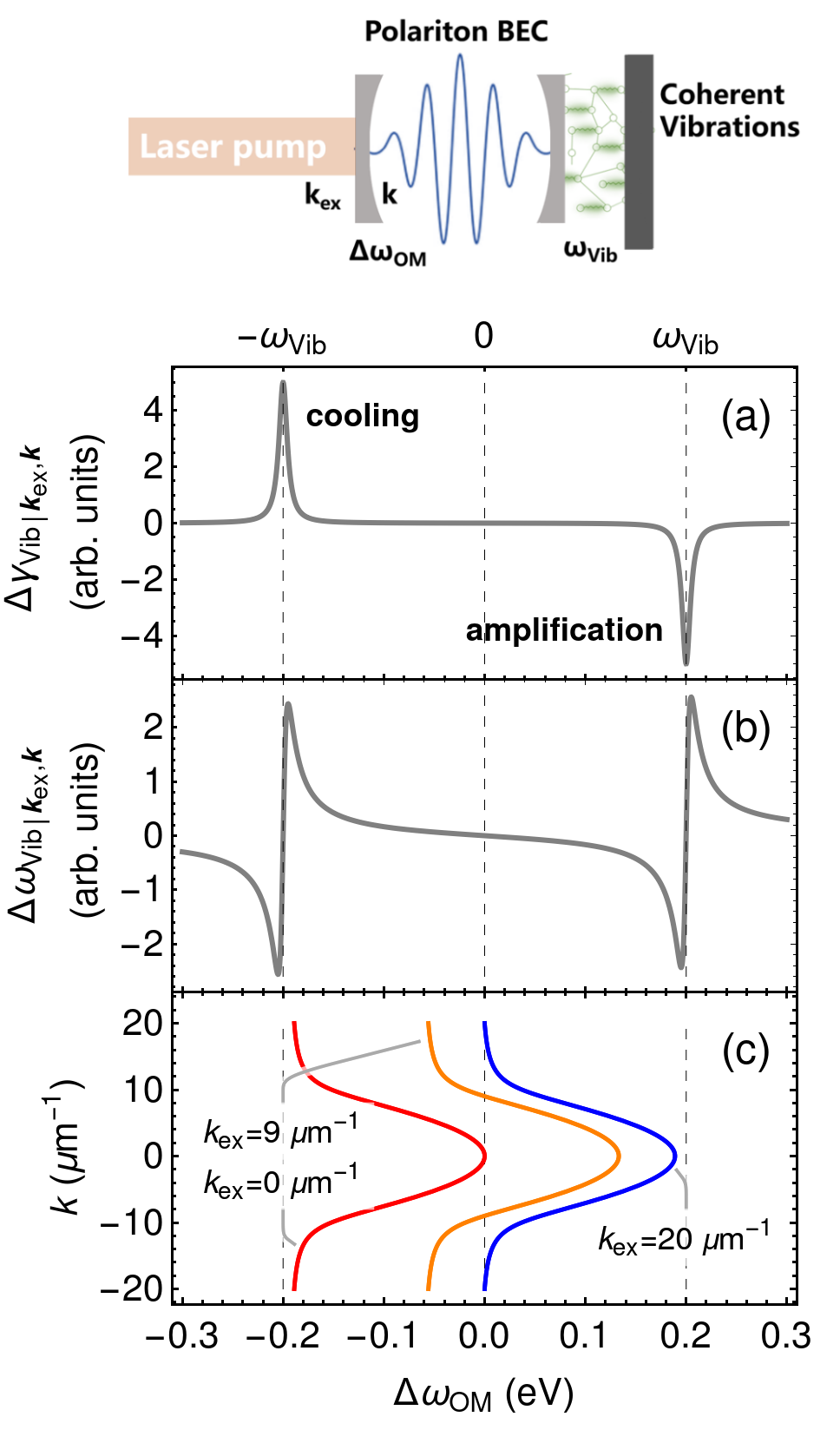}
\caption{
Optomechanically-induced damping $\Delta \gamma_{{\rm Vib}|{{\bf k}_{\rm ex},{\bf k}}}$ -- (a) and frequency shift $\Delta \omega_{{\rm Vib}|{{\bf k}_{\rm ex},{\bf k}}}$ -- (b) of the vibrational mode at $\Delta \omega_{\rm OM}=\Delta \omega_{{\bf k}_{\rm ex}{\bf k}} = \omega_{{\rm Low}|{\bf k}_{\rm ex}} - \omega_{{\rm Low}|{\bf k}}$. The momentum dependence of the optomechanical detuning $\Delta \omega_{{\bf k}_{\rm ex}{\bf k}}$-- (c) at different ${\bf k}_{\rm ex}$ of the laser drive.
Vertical dashed lines correspond to the red-detuned $\Delta \omega_{{\bf k}_{\rm ex}{\bf k}} = -\omega_{\rm Vib}$; zero-detuned $\Delta \omega_{{\bf k}_{\rm ex}{\bf k}} = 0$; and blue-detuned resonant interactions $\Delta \omega_{{\bf k}_{\rm ex}{\bf k}} = +\omega_{\rm Vib}$ . Here we assume momentum-independent values of the interaction constant $g_{{\bf k}_{\rm ex}{\bf k}}$ and dephasing rate of resonantly-driven polariton states $\Gamma_{{\rm low}|{\bf k_{ex}}}$ according to parameters in Table~II.
}
    \label{fig:susceptibility_1D_to_2D}
\end{figure}

\begin{figure*}
\includegraphics[width=1\linewidth]{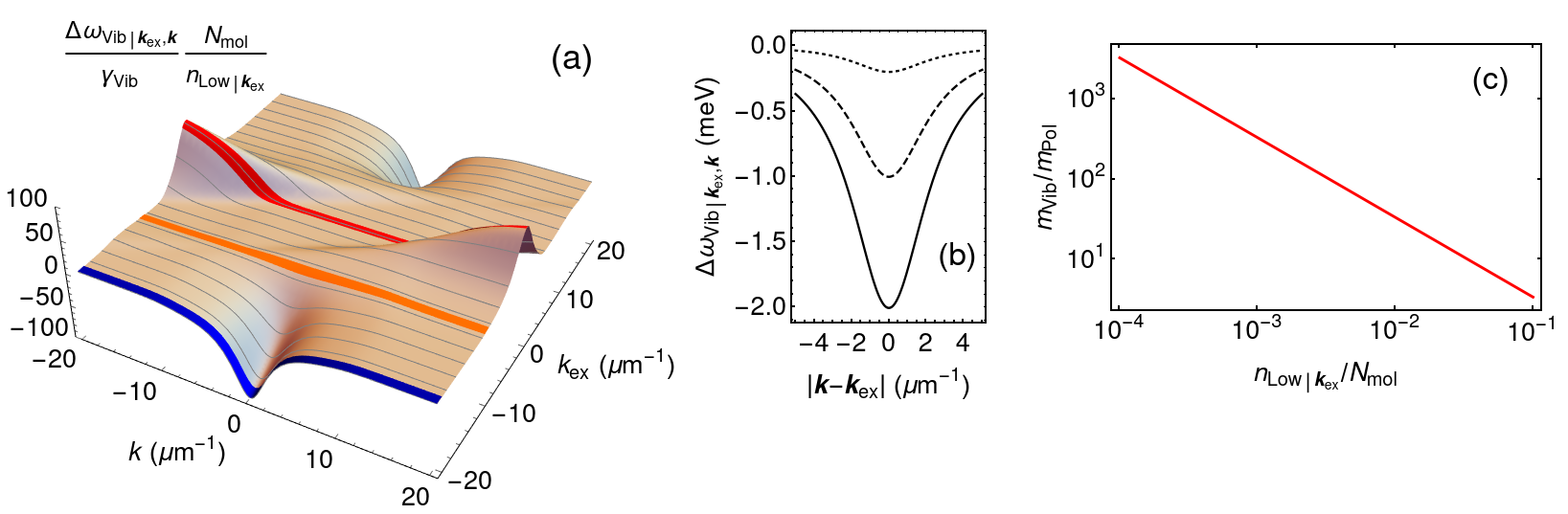}
\caption{
Normalized frequency shift of the vibrational mode as a function of the momentum of both the resonantly-driven polariton state ${\bf k}_{\rm ex}$ and the polariton condensate ${\bf k}$ -- (a).
Blue, orange, and red lines correspond to the cross-sections taken for the certain laser-driven states ${\bf k}_{\rm ex}$ in accordance with Fig.~\ref{fig:susceptibility_1D_to_2D}(c).
Energy change of the vibrational mode as a function of polariton momenta of the condensate -- (b).
Dotted, dashed and solid lines represent the energy change obtained at the following occupation of the resonantly-driven state $n_{{\rm Low}|{{\bf k}_{\rm ex}}}$: $10^5$; $5\cdot10^5$; and $10^6$ respectively.
The effective mass ratio between the vibrational mode $m_{\rm Vib} = \hbar/2\alpha_{\rm Vib}$ and lower polaritons $m_{\rm Pol} = \hbar/2\alpha_{\rm Pol}$ -- (c).
We use the following main parameters: $\Lambda^{2} = 1$, $\omega_{\rm Vib}=200$~$\rm{meV}$, $N_{\rm mol}=10^8$, $\gamma_{\rm Exc}=10^{-5}$~$\rm{eV}$, $\Gamma_{\rm Exc} = 10^{-2}$~$\rm{eV}$, $\gamma_{\rm Vib} = 2\cdot 10^{-3}$~$\rm{eV}$, $\gamma_{\rm {Cav}|{\bf{k}}}=2.5\cdot10^{-3}$~$\rm{eV}$, see Table~II.
}
    \label{fig:effective_vib_mass}
\end{figure*}

Given the typical frequencies of high-energy vibrational modes ($\hbar\omega_{\rm Vib}\sim100~{\rm meV}$) and the cavity linewidth of exciton-polariton microcavities ($\hbar\gamma_{{\rm Cav}|{\bf k}}\sim1~{\rm meV}$) the system stays in a very well resolved sideband regime ($\Gamma_{{\rm Low}|{\bf k}}\simeq\gamma_{{\rm Cav}|{\bf k}}\ll\omega_{\rm Vib}$). The character of the linearized optomechanical interaction in Eq.~(\ref{H_bright}) depends on the laser detuning with respect to the polariton state at the condensate $\Delta\omega_{\rm OM}=\omega_{{\rm Low}|{\bf k}_{\rm ex}}-\omega_{{\rm Low}|{\bf k}}$. 
In the sideband-resolved regime one can distinguish between three distinct resonant configurations: (1) red-detuned ($\Delta\omega_{\rm OM}=-\omega_{\rm Vib}$); (2) blue-detuned ($\Delta\omega_{\rm OM}=+\omega_{\rm Vib}$); and (3) zero-detuned ($\Delta\omega_{\rm OM}=0$) interaction, as shown in Fig.~\ref{fig:susceptibility_1D_to_2D}.

In the first, red-detuned configuration the main contribution comes from the resonant terms in the form $\sim\hbar g_{{\bf k}_{\rm ex}{\bf k}}\sqrt{n_{{\rm Low}{\bf k}_{\rm ex}}} ( \hat{S}_{{\rm Low}|{\bf k}}^\dag \hat{C}_{{\rm Vib}|{\bf k}-{\bf k}_{\rm ex}} + \hat{S}_{{\rm Low}|{\bf k}} \hat{C}_{{\rm Vib}|{\bf k}-{\bf k}_{\rm ex}}^\dag)
$. Known as a ``beam-splitter'' type interaction~\cite{scully1997quantum} this Hamiltonian results in the energy exchange between polariton BEC and the collective vibrational mode. In sideband resolved zero-dimensional optomechanics this interaction imposes a positive optomechanical damping $\Delta \gamma_{{\rm Vib}|{{\bf k}_{\rm ex},{\bf k}}}>0$ that counteracts mechanical motion leading to the cooling effect~\cite{aspelmeyer2014cavity}. Fig.~\ref{fig:susceptibility_1D_to_2D}a illustrates the additional damping of the coherent vibrational mode that results from the red-detuned optomechanical interaction.

The blue-detuned optomechanical configuration characterized by fast-rotating terms which are dominant in the interaction $\sim\hbar g_{{\bf k}_{\rm ex}{\bf k}}\sqrt{n_{{\rm Low}{\bf k}_{\rm ex}}} ( \hat{S}_{{\rm Low}|{\bf k}}^\dag \hat{C}_{{\rm Vib}|{\bf k}_{\rm ex}-{\bf k}}^\dag + \hat{S}_{{\rm Low}|{\bf k}} \hat{C}_{{\rm Vib}|{\bf k}_{\rm ex}-{\bf k}})
$. The creation (or annihilation) of collective vibrational quanta of nuclei motion alongside with polariton quanta at the BEC essentially represents ``two-mode squeezing'' interaction. In conventional optomechanical systems, interaction of this type leads to parametric amplification of mechanical motion~\cite{clerk2010introduction}. In our case, we have polariton BEC instead of a bare cavity mode, therefore, driven by the blue-detuned laser drive this term leads to exponential growth of of both, the amplitude of the collective vibrational mode and the BEC density, and gives rise to strong correlations between the two. For vibrational degrees of freedom it results in the substantial decrease of the damping $\gamma_{\rm Vib}$, which is evident in Fig.~\ref{fig:susceptibility_1D_to_2D}a as the polariton optomechanical antidamping effect $\Delta \gamma_{{\rm Vib}|{{\bf k}_{\rm ex},{\bf k}}}<0$, In the next section, we solely focus on this regime discussing parametric amplification effect toward vibrational condensation.

The frequency change of the collective vibrational mode is another important consequence of the optomechanical interaction. According to Eq.~(\ref{optomechanical_shift}), the frequency shift depends heavily on the optomechanical detuning and the polariton decay rate $\Gamma_{{\rm Low}|{\bf k}}$. For practical microcavity-based polariton systems and high-frequency vibrational modes we operate in the well-defined sideband resolved regime as shown in Fig.~\ref{fig:susceptibility_1D_to_2D}b. The decrease or increase in the vibrational frequency around the resonant detuning $\Delta\omega_{\rm OM}=\pm\omega_{\rm Vib}$ induced by the optomechanical interaction can be interpreted as vibrational spring softening or hardening, respectively. 
In the Doppler regime ($\gamma_{{\rm Cav}|{\bf k}}>\omega_{\rm Vib}$) it is known as an optical spring effect~\cite{aspelmeyer2014cavity}.

So far, we have not addressed the direct implications of the two-dimensional nature of polariton states. Fig.~\ref{fig:susceptibility_1D_to_2D}b showcases an effect that is zero-dimensional~\cite{aspelmeyer2014cavity} neglecting certain resonant conditions for the momenta ${\bf k}_{\rm ex}$ and $\bf k$ in the dispersion relation of polariton states (see Fig.~\ref{fig:dispersion}). In fact, the optomechanical detuning $\Delta\omega_{\rm OM}$ inherits genuine k-dependence from lower polariton states. In Fig.~\ref{fig:susceptibility_1D_to_2D}c we represent only three dispersion relations tied to particular momenta out of an infinite number of possible realizations in the full two-dimensional $\{E,k\}$ - phase space. Here, one can see that the range of possible optomechanical detuning values $\Delta \omega_{\rm OM}$ is bound by the dispersion relation of polariton states. Therefore, the actual frequency shift of the vibrational mode due to the polariton optomechanical interaction is determined by the convolution of the zero-dimensional change in Fig.~\ref{fig:susceptibility_1D_to_2D}b and detuning dependencies represented in Fig.~\ref{fig:susceptibility_1D_to_2D}c.

The frequency shift of the vibrational mode $\Delta \omega_{{\rm Vib}|{{\bf k}_{\rm ex},{\bf k}}}$ in two-dimensional momentum space is depicted in Fig.~\ref{fig:effective_vib_mass}a. Here, we focus on three specific cross-sections along the polariton momentum $\hbar {\bf k}$, with in-plane momentum of the laser drive $\hbar {\bf k}_{\rm ex}$ held constant, linking the comprehensive 2D polariton landscape to the simplified 0D optomechanical picture represented in Fig.~\ref{fig:susceptibility_1D_to_2D}.
The coherent interaction gives rise to the dispersion relation for the vibrational mode. Being dispersionless otherwise, molecular vibrations acquire polariton-like dispersion relation conditioned on the optomechanical detuning. Importantly, the vibrational dispersion is nearly parabolic at around  ${\bf k}={\bf 0}$ in the resonant blue-detuned configuration $\Delta \omega_{{\bf k}_{\rm ex}{\bf 0}} \approx +\omega_{\rm Vib}$, such that we can approximate Eq.~(\ref{optomechanical_shift}) by
\begin{equation}
\Delta \omega_{{\rm Vib}|{{\bf k}_{\rm ex},{\bf k}}}
\approx
\Delta \omega_{{\rm Vib}|{{\bf k}_{\rm ex},{\bf 0}}}
+
\alpha_{\rm Vib} {\bf k}^2,
\end{equation}
where
\begin{multline} \label{vib dispersion}
\alpha_{\rm Vib} 
= 
\alpha_{\rm Pol}
|g_{{\bf k}_{\rm ex}{\bf k}={\bf 0}}|^2 n_{{\rm Low}|{{\bf k}_{\rm ex}}}
\\
\frac
{
\left( \omega_{\rm Vib} - \Delta \omega_{{\bf k}_{\rm ex}{\bf k}={\bf 0}} \right)^2 - 
\left( \Gamma_{{\rm Low}|{\bf k}={\bf 0}}/2\right)^2
}{
\left[\left( \omega_{\rm Vib} - \Delta \omega_{{\bf k}_{\rm ex}{\bf k}={\bf 0}} \right)^2 + 
\left( \Gamma_{{\rm Low}|{\bf k}={\bf 0}}/2\right)^2 \right]^2
}  .
\end{multline}

The effective mass of coherent molecular vibrations can be calculated as $m_{\rm Vib} = \hbar/2\alpha_{\rm Vib}$. The effective mass remains positive unless $|\Delta \omega_{{\bf k}_{\rm ex}{\bf k}={\bf 0}} - \omega_{\rm Vib}| < \Gamma_{{\rm Low}|{\bf k}={\bf 0}}/2$. 
From Eq.~(\ref{vib dispersion}), the effective mass of coherent vibrations, $m_{\rm Vib}$, can be directly compared to the effective mass of polaritons, $m_{\rm Pol} = \hbar/2\alpha_{\rm Pol}$. This ratio depends on optomechanical coupling strength and intensity of the laser drive included here in $ n_{{\rm Low}|{{\bf k}_{\rm ex}}}$ parameter. 
Fig.~\ref{fig:effective_vib_mass} shows the ratio between effective masses of the vibrational and polariton subsystems as the function of excitation density $ n_{{\rm Low}|{{\bf k}_{\rm ex}}}/N_{\rm mol}$ resonantly driven by the laser. In analogy to polariton states, the positive effective mass has direct implications for nonequilibrium Bose-Einstein condensation~\cite{deng2010exciton}, which we now propose for molecular vibrations~\cite{shishkov2023mapping}. Control over the optomechanical interaction can be used to manipulate the dispersion of coherent vibrational and phonon states. 
A similar analysis can be applied to red-detuned optomechanical interactions, which may result in a negative effective mass for molecular vibrations. 
The negative effective mass, in turn, allows for soliton formation, enabled by self-focusing phenomena in cases of repulsive interactions between polaritons~\cite{sich2012observation}. When periodic potentials are applied, anomalous dispersion can also lead to a negative effective mass, as has been demonstrated experimentally in ultra-cold Rb atoms~\cite{eiermann2004bright} and proposed for exciton-polariton systems~\cite{teklu2016non}. The potential of the negative effective mass of collective vibrational states coupled to polariton BECs has yet to be explored.

In the next Section, we study transition from thermal polariton and vibrational states to their condensates at the resonant  optomechanical interaction.

\section{Results and discussion} \label{sec:consequences}

The coherent optomechanical picture developed above describes the regime of well-established polariton condensation, where we can neglect manifolds of the dark states. However, dark states play a significant role in the BEC formation. In order to follow the condensation process, we turn to the Hamiltonian described in Eq.~(\ref{FullHamiltonian_polaritons}) and proceed to derive the dynamics of the average occupation numbers for all states involved, as illustrated in Fig.~\ref{fig:Layout}. This includes the average number of lower polaritons with the wavevector $\bf k$, $n_{{\rm Pol}|{\bf k}}=\langle \hat s_{{\rm Low}|{\bf k}}^\dag \hat s_{{\rm Low}|{\bf k}} \rangle$, the average number of dark excitons per molecule, $n_{\rm Exc_D} = \langle \hat n_{\rm Exc_D} \rangle$, average number of the dark vibrations per molecule, $n_{\rm Vib_D} = \langle \hat n_{\rm Vib_D} \rangle$, bright excitons with the wavevector ${\bf k}_{\rm ex}$, $n_{{\rm Exc}|{\bf k}_{\rm ex}}=\langle \hat s_{{\rm Low}|{\bf k}_{\rm ex}}^\dag \hat s_{{\rm Low}|{\bf k}_{\rm ex}} \rangle$ and the average number of bright vibrations with wave vector $\bf q$, $n_{{\rm Vib}|{\bf q}} = \langle \hat c_{{\rm Vib}|{\bf q}}^\dag \hat c_{{\rm Vib}|{\bf q}} \rangle$,
\begin{multline} \label{NumberOfBrightExcitons+flow}
\frac{d n_{{\rm Exc}|{\bf k}_{\rm ex}}}{dt} 
=  
-  \gamma_{{\rm Low}|{\bf k}_{\rm ex}} ( n_{{\rm Exc}|{\bf k}_{\rm ex}} - \varkappa_{\rm Pump} )
+
\\
\gamma_{\rm Exc}^{\rm B-D} ( n_{\rm Exc_D} - n_{{\rm Exc}|{\bf k}_{\rm ex}} ) 
-
\sum_{\bf k} J_{\bf k}^{({\rm Bright})},
\end{multline}
\begin{multline} \label{NumberOfDarkExcitons+flow}
\frac{d n_{\rm Exc_D}}{dt} 
=  
-  \gamma_{\rm Exc} n_{\rm Exc_D}
+
\\
\frac{ \gamma_{\rm Exc}^{\rm B-D} }{ N_{\rm mol}} ( n_{{\rm Exc}|{\bf k}_{\rm ex}} - n_{\rm Exc_D} ) 
-
\frac{1}{N_{\rm mol}} \sum_{\bf k} J_{\bf k}^{({\rm Dark})},
\end{multline}
\begin{multline} \label{NumberOfLowerPolaritonsH+flow}
\frac{d n_{{\rm Pol}|{\bf k}}}{dt} 
=  
-  \gamma_{{\rm Low}|{\bf k}} n_{{\rm Pol}|{\bf k}}
+ 
J_{\bf k}^{({\rm Full})}
+
\\
\sum_{\bf k'} \left\{ \gamma _{\rm therm}^{\bf kk'}\left( n_{{\rm Pol}|{\bf k}} + 1 \right){n_{{\rm Pol}|{\bf k'}}} 
 - 
\right.
\\ 
\left.
\gamma _{{\rm{therm}}}^{{\bf{k'k}}}\left( n_{{\rm Pol}|{\bf k'}} + 1 \right){n_{{\rm Pol}|{\bf k}}} \right\} ,
\end{multline}
\begin{multline} \label{NumberOfDarkmolecular vibrations+flow}
\frac{d n_{\rm Vib_D}}{dt} 
=  
-  \gamma_{\rm Vib} \left( n_{\rm Vib_D} - n_{\rm Vib}^{\rm th}  \right) 
+
\\
\frac{\gamma_{\rm Vib}^{\rm B-D} }{ N_{\rm mol}} \sum_{\bf k} 
\left( n_{{\rm Vib}|{\bf k_{\rm ex}}-{\bf k}} - n_{\rm Vib_D} \right)
+ 
\frac{1}{N_{\rm mol}} \sum_{\bf k} J_{\bf k}^{({\rm Dark})},
\end{multline}
\begin{multline} \label{NumberOfBrightmolecular vibrations+flow}
\frac{d n_{{\rm Vib}|{\bf k_{\rm ex}}-{\bf k}}}{dt} 
=  
-  \gamma_{\rm Vib} \left( n_{{\rm Vib}|{\bf k_{\rm ex}}-{\bf k}} - n_{{\rm Vib}}^{\rm th}  \right)
+
\\
\gamma_{\rm Vib}^{\rm B-D} 
\left( n_{\rm Vib_D} - n_{{\rm Vib}|{\bf k_{\rm ex}}-{\bf k}} \right)
+ 
J_{\bf k}^{({\rm Bright})},
\end{multline}
where we approximated $\hat n_{\rm Exc_D} \approx N_{\rm mol}^{-1} \sum_{j=1}^{N_{\rm mol}} \hat S_{{\rm Exc}j}^\dag \hat S_{{\rm Exc}j} $, and $\hat n_{\rm Vib_D} \approx N_{\rm mol}^{-1} \sum_{j=1}^{N_{\rm mol}}  \mathcal{\hat B}_{{\rm Vib}j}^\dag \mathcal{\hat B}_{{\rm Vib}j}$ with $\hat S_{{\rm Exc}j}$ being the operator of vibrationally dressed excitons and $\hat B_{{\rm Vib}j}$ being the operator of dressed molecular vibrations for the $j$th molecule defined in Appendix A by Eq.(\ref{dressed_excitons})--(\ref{dressed_vibrons}).
We introduce the energy flows for the bright and dark states:
\begin{multline}
J_{\bf k}^{({\rm Full})}
=
-i \cos\varphi_{\bf k}
\\
\sum_{j=1}^{N_{\rm mol}} \frac{\Lambda \Omega_R}{\sqrt{N_{\rm mol}}}
\langle \hat S^\dag_{{\rm Exc}j} \hat s_{{\rm Low}|{\bf k}} \hat B_{{\rm Vib}j} \rangle 
e^{i{\bf k}{\bf r}_j}
+
h.c.,
\end{multline}
\begin{multline} \label{J_bright}
J_{\bf k}^{({\rm Bright})}
=
i \cos\varphi_{\bf k} \sin\varphi_{{\bf k}_{\rm ex}}
\\
\frac{\Lambda \Omega_R}{\sqrt{N_{\rm mol}}}
\langle 
\hat s^\dag_{{\rm Low}|{\bf k}_{\rm ex}} \hat s_{{\rm Low}|{\bf k}} \hat c_{{\rm Vib}|{\bf k}_{\rm ex}{\bf -k}} 
\rangle
+
h.c.,
\end{multline}
\begin{equation} \label{J_dark}
J_{\bf k}^{({\rm Dark})}
=
J_{\bf k}^{({\rm Full})} - J_{\bf k}^{({\rm Bright})}.
\end{equation}

The resultant Eq.~(\ref{NumberOfBrightExcitons+flow})--(\ref{NumberOfBrightmolecular vibrations+flow}) describe the dynamics of the average occupation number in the system. Fig.~\ref{fig:Layout} illustrates the energy flow among distinct subsystems, including bright and dark excitons, molecular vibrations, and polaritons at the state with $\hbar\bf{k}$ momentum.  In densely packed molecular systems and semiconductor materials at room-temperature, the dephasing rate of excitons, $\Gamma_{\rm Exc}$, is significantly faster than any other relaxation processes involved, such as $\gamma_{\rm Exc}$, $\gamma_{{\rm Low}|{\bf k}}$, or $\gamma_{{\rm Vib}}$~(see Table~\ref{table: parameters}). Consequently, the flows $J_{\bf k}^{({\rm Bright})}$ and $J_{\bf k}^{({\rm Dark})}$ almost instantly adjust to the current occupation numbers of the excitonic, vibrational, and polaritonic states. Thus, we adiabatically exclude them from the dynamics and obtain the following

\begin{multline} \label{NumberOfBrightExcitons}
\frac{d n_{\rm Exc|{\bf k}_{\rm ex}}}{dt} 
=  
-  \gamma_{\rm Exc} ( n_{\rm Exc|{\bf k}_{\rm ex}} - \varkappa_{\rm Pump} )
+ 
\gamma_{\rm Exc}^{\rm B-D} ( n_{\rm Exc_{D}} - 
\\
n_{{\rm Exc}|{\bf k}_{\rm ex}} ) 
-
\sum_{\bf k} \frac{\tilde G_{{\bf k}_{\rm ex}{\bf k}}}{N_{\rm mol}} 
\left[
n_{{\rm Exc}|{\bf k}_{\rm ex}} \left(n_{{\rm Pol}|{\bf k}} + 1 \right) 
+
\right.
\\
\left.
n_{{\rm Vib}|{\bf k_{\rm ex}}-{\bf k}} \left( n_{{\rm Pol}|{\bf k}_{\rm ex}} - n_{{\rm Exc}|{\bf k}_{\rm ex}} \right) 
\right],
\end{multline}
\begin{multline} \label{NumberOfDarkExcitons}
\frac{d n_{\rm Exc_{D}}}{dt} 
=  
-  \gamma_{\rm Exc} n_{\rm Exc_{D}}
+ \frac{ \gamma_{\rm Exc}^{\rm B-D} }{ N_{\rm mol}} ( n_{{\rm Exc}|{\bf k}_{\rm ex}} - n_{\rm Exc_{D}} ) 
-
\\
\sum_{\bf k} \frac{G_{\bf k}}{N_{\rm mol}} 
\left[
n_{\rm Exc_{D}} \left(n_{{\rm Pol}|{\bf k}} + 1 \right) 
+
n_{\rm Vib_{D}} \left( n_{\rm Exc_{D}} - n_{{\rm Pol}|{\bf k}} \right) 
\right],
\end{multline}
\begin{multline} \label{NumberOfLowerPolaritons}
\frac{d n_{{\rm Pol}|{\bf k}}}{dt} 
=  
-  \gamma_{{\rm Pol}|{\bf k}} n_{{\rm Pol}|{\bf k}}
+ 
G_{\bf k}
\left[
n_{\rm Exc_{D}} \left(n_{{\rm Pol}|{\bf k}} + 1 \right) 
+
 \right.
\\
\left.
n_{\rm Vib_{D}} \left( n_{\rm Exc_{D}} - n_{{\rm Pol}|{\bf k}} \right) 
\right]
+ 
\frac{ \tilde G_{{\bf k}_{\rm ex}{\bf k}} }{ N_{\rm mol} } 
\left[
n_{\rm Exc|\bf {k_{ex}}} \left(n_{{\rm Pol}|{\bf k}} + 
\right.
\right.
\\
\left.
\left.
1 \right) 
+
n_{{\rm Vib}|{\bf k_{\rm ex}}-{\bf k}} \left( n_{{\rm Exc}|{\bf k}_{\rm ex}} - n_{{\rm Pol}|{\bf k}} \right) 
\right]
 + 
\sum_{\bf k'} \left\{ \gamma _{\rm therm}^{\bf kk'}
\right.
\\
\left.
\left( n_{{\rm Pol}|{\bf k}} + 
1 \right){n_{{\rm Pol}|{\bf k'}}} 
 - 
\gamma _{{\rm{therm}}}^{{\bf{k'k}}}\left( n_{{\rm Pol}|{\bf k'}} + 1 \right){n_{{\rm Pol}|{\bf k}}} \right\} ,
\end{multline}
\begin{multline} \label{NumberOfDarkmolecular vibrations}
\frac{d n_{\rm Vib_{D}}}{dt} 
=  
-  \gamma_{\rm Vib} \left( n_{\rm Vib_{D}} - n_{\rm Vib}^{\rm th}  \right) 
+ 
\frac{ \gamma_{\rm Vib}^{\rm B-D} }{ N_{\rm mol}}
\sum_{\bf k} 
\left(
\right.
\\
\left.
 n_{{\rm Vib}|{\bf k}_{\rm ex}-{\bf k}} - 
n_{\rm Vib_{D}} \right)
+ 
\sum_{\bf k}
 \frac{ G_{\bf k}}{N_{\rm mol}}  
\left[
n_{\rm Exc_{D}} \left(n_{{\rm Pol}|{\bf k}} + 
1 \right) 
+
\right.
\\
\left.
n_{\rm Vib_{D}} \left( n_{\rm Exc_{D}} - n_{{\rm Pol}|{\bf k}} \right) 
\right],
\end{multline}
\begin{multline} \label{NumberOfBrightmolecular vibrations}
\frac{d n_{{\rm Vib}|{\bf k}_{\rm ex}-{\bf k}}}{dt} 
=  
-  \gamma_{\rm Vib} \left( n_{{\rm Vib}|{\bf k}_{\rm ex}-{\bf k}} - n_{{\rm vib}}^{\rm th}  \right)
+
 \gamma_{\rm Vib}^{\rm B-D} 
\\
\left( n_{\rm Vib_{D}} - n_{{\rm Vib}|{\bf k}_{\rm ex}-{\bf k}} \right)
+ 
\frac{\tilde G_{{\bf k}_{\rm ex}{\bf k}}}{N_{\rm mol}} 
\left[
n_{\rm Exc|{\bf k}_{\rm ex}} \left(n_{{\rm Pol}|{\bf k}} + 1 \right) 
+
 \right.
\\
\left.
n_{{\rm Vib}|{\bf k_{\rm ex}}-{\bf k}} \left( n_{\rm Exc|{\bf k}_{\rm ex}} - n_{{\rm Pol}|{\bf k}} \right) 
\right],
\end{multline}
where we denote 
\begin{equation}
\omega_{{\rm Pol}|{\bf k}} = \omega_{{\rm Low}|{\bf k}},  
\end{equation}
\begin{equation}
\gamma_{{\rm Pol}|{\bf k}} = \gamma_{{\rm Low}|{\bf k}},
\end{equation}
\begin{equation}\label{optomechanical_constant_2}
\tilde G_{{\bf k}_{\rm ex}{\bf k}}
=
\frac
{ 
\Lambda^2 \Omega_R^2 \Gamma_{\rm Exc} \cos^2\varphi_{\bf k} \sin^2 \varphi_{{\bf k}_{\rm ex}}
}{ 
(\omega_{\rm Exc}-\omega_{{\rm Pol}|{\bf k}}-\omega_{{\rm Vib}})^2 + (\Gamma_{\rm Exc}/2)^2 
},
\end{equation}
\begin{equation}\label{optomechanical_constant}
 G_{\bf k}
=
\frac{ \Lambda^2 \Omega_R^2 \Gamma_{\rm Exc} \cos^2\varphi_{\bf k}}{ (\omega_{\rm Exc}-\omega_{{\rm Pol}|{\bf k}}-\omega_{{\rm Vib}})^2 + (\Gamma_{\rm Exc}/2)^2 }.
\end{equation}

Optomechanical constants $G_{\bf k}$ and $\tilde G_{{\bf k}_{\rm ex}{\bf k}}$ play the central role in polariton condensation and generation of macroscopic vibrational states.
In the blue-detuned configuration, it defines a net vibrational amplification parameter $\tilde G_{{\bf k}_{\rm ex}{\bf k}}n_{{\rm Low}|{\bf k}_{\rm ex}}$, which is similar to the previously mentioned optomechanical antidamping rate, $\Delta \gamma_{{\rm Vib}|{{\bf k}_{\rm ex},{\bf k}}}$, as given by Eq.~(\ref{optomechanical_constant}). The primary difference is that the latter includes $\Gamma_{{\rm Low}|{\bf k}}$ instead of $\Gamma_{\rm Exc}$.
This is because our optomechanical setup is based on the continuous (or~\textit{quasi-continuous}) laser drive, locking the phase and excluding the dephasing rate of the polariton state from optomechanical damping.
Below, we focus on the blue-detuned optomechanical interaction, aligning the laser drive in resonance with high-${\it k}$ polariton states that feature a large wave vector ${\bf k}_{\rm ex}$, as depicted in Fig.~\ref{fig:dispersion}. We exclude the upper polariton states from consideration because they are far from the resonant excitation. Indicated by $\varphi_{{\bf k}_{\rm ex}} \approx \pi/2$, this approximation implies $\omega_{{\rm Up}|{\bf k}_{\rm ex}} \approx \omega_{{\rm Cav}|{\bf k}_{\rm ex}}$ and $\omega_{{\rm Low}|{\bf k}_{\rm ex}} \approx \omega_{\rm Exc}$, with $\gamma_{{\rm Low}|{\bf k}_{\rm ex}} \approx \gamma_{\rm Exc}$. Consequently, $\tilde G_{{\bf k}_{\rm ex}{\bf k}} \approx G_{\bf k}$, which is applied in the simulations. Additionally, we consider a cavity with a quadratic dispersion relation.

\begin{equation}
\omega_{{\rm Cav}|{\bf k}} 
= 
\omega_{{\rm Cav}|{\bf k}={\bf 0}}
+
\alpha_{\rm Cav} {\bf k}^2,
\end{equation}
where $\omega_{\rm Exc} - \omega_{{\rm Pol}|{\bf k}={\bf 0}} \approx \omega_{\rm Vib}$, such that from Eq.~(\ref{FrequenciesOfLowerPolaritons}) it becomes
\begin{equation} \label{polariton dispersion}
\omega_{{\rm Pol}|{\bf k}} 
= 
\omega_{{\rm Pol}|{\bf k}={\bf 0}}
+
\alpha_{\rm Pol} {\bf k}^2,
\end{equation}
with
\begin{equation}
\alpha_{\rm Pol}
=
\alpha_{\rm Cav}
\frac{
\omega_{\rm Vib}^2
}{
\omega_{\rm Vib}^2 + \Omega_R^2
}.
\end{equation}
Table~\ref{table: parameters} provides the parameters that we use in the calculations below.

\subsection{Polariton condensation}

\begin{figure}
\includegraphics[width=1\linewidth]{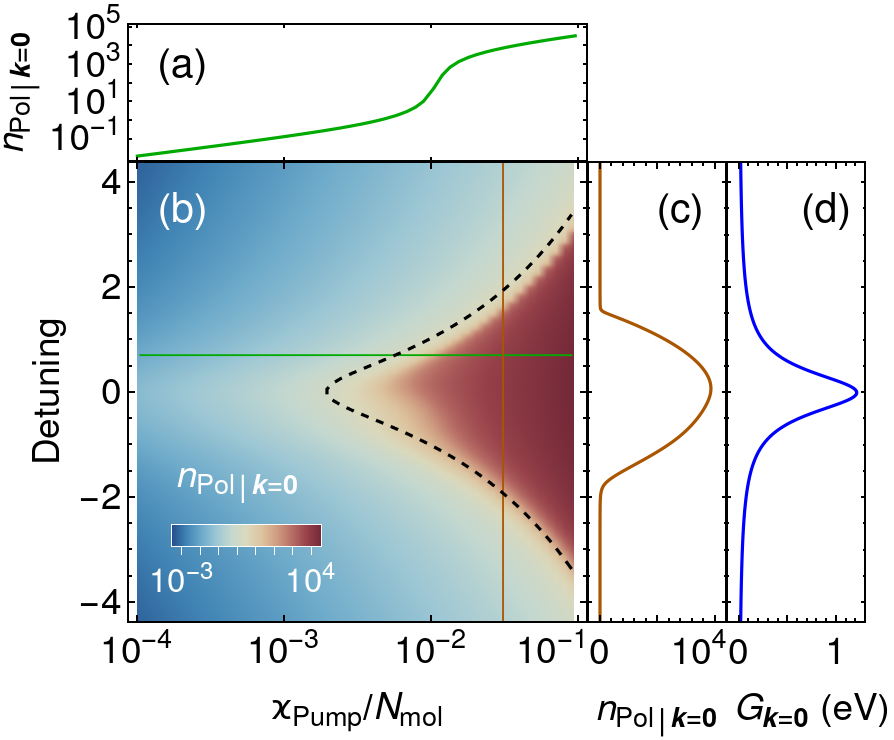}
\caption{
Polariton occupation at the ground state as a function of pumping parameter -- (a), taken at the particular value of the normalized detuning $(\omega_{{\rm Pol}|{\bf k}={\bf 0}}+\omega_{\rm Vib}-\omega_{\rm Exc})/\Gamma_{\rm Exc}$ (green line) from the density plot -- (b). The black dashed line shows the analytical threshold value derived in Ref.~\cite{shishkov2023mapping}. Polariton occupation at the ground state as a function of the detuning taken at the fixed pumping conditions (brown line) -- (c). Dependence of the optomechanical constant $G_{{\bf k}={\bf 0}}$~[Eq.~(\ref{optomechanical_constant})] as a function of the normalized detuning.
The parameters are the following $\Lambda^{2} = 1$, $N_{\rm mol}=10^8$, $\gamma_{\rm Exc}=10^{-5}$~$\rm{eV}$, $\gamma_{\rm Exc}^{\rm B-D} = \Gamma_{\rm Exc} = 10^{-2}$~$\rm{eV}$,
$\gamma_{\rm Vib}^{\rm B-D} = \Gamma_{\rm Vib}$, 
$\gamma_{{\rm{therm}}}^{\bf kk'}= 10^{-5}$~$\rm eV$ for $|{\bf k}| < |{\bf k'}|$ and $T=290$~$\rm{K}$,
$\gamma_{{\rm Pol}|{\bf k}}\approx\gamma_{\rm {Cav}|{\bf{k}}}=2.5
\cdot10^{-3}$~$\rm{eV}$, $S = 500$~$\mu{\rm m^2}$, see Table~II
}
    \label{fig:av_pol_occ_num}
\end{figure}

Here we present the results of numerical simulations for the steady-state density of polaritons at the ground state. Fig.~\ref{fig:av_pol_occ_num}b is the color plot of the polariton density as function of the normalized detuning $\Delta\omega = (\omega_{{\rm Pol}|{\bf k}={\bf 0}}+\omega_{\rm Vib}-\omega_{\rm Exc})/\Gamma_{\rm Exc}$ and pumping $\varkappa_{\rm Pump}/N_{\rm mol}$ parameters. Polaritons exhibit superlinear increase in density when the pump exceeds a certain threshold value ($P_{th}$) indicating the onset of the BEC formation. The threshold behaviour is illustrated in the cross-section taken along the x-axis in Fig.~\ref{fig:av_pol_occ_num}a. In Ref.~\cite{shishkov2023mapping} we provide further evidence of polariton condensation, which includes Bose--Einstein distributions and off-diagonal long range order. The central role of the strong exciton-vibration interaction in the polariton condensation can be seen from the detuning dependence in Fig.~\ref{fig:av_pol_occ_num}c, taken as the cross-section along y-axis of the color plot. Under the resonant condition $\omega_{{\rm Pol}|{\bf k}={\bf 0}}+\omega_{\rm Vib}-\omega_{\rm Exc}=0$ underlying polariton optomechanical interaction, this mechanism efficiently channels the population from exciton states to the ground polariton states. Indeed, the detuning dependence of polariton density is fully consistent with the observed resonant properties of the optomechanical coupling strength~$G_{\bf k}$, as shown in~Fig.~\ref{fig:av_pol_occ_num}d. Note, the width of the resonance is quite broad comparing narowline vibrational modes typically accessible in Raman spectroscopy. This is a feature of the polariton optomechanical interaction that encompasses exciton states. The large dephasing rate $\Gamma_{\rm Exc}$ makes it broadband, within the exciton linewidth. 


These results are in good agreement with experimental observations~\cite{zasedatelev2021single, zasedatelev2019room}. Early studies reported similar detuning dependences in the total polariton density in microcavity structures below condensation threshold~\cite{coles2011vibrationally,somaschi2011ultrafast}. Recent experiments discover  correlations in the total polariton density at the BEC with the vibrational resonances following the detuning parameter controlled by the cut-off frequency of the cavity and photon energy of the laser drive~\cite{zasedatelev2019room}. 
Additionally, resonant nature of the optomechanical interaction leads to sharp dependence in condensation threshold indicated by the dashed line in Fig.~\ref{fig:av_pol_occ_num}b. The lowest threshold for polariton BEC occurs when $\omega_{{\rm Pol}|{\bf k}={\bf 0}}+\omega_{\rm Vib}-\omega_{\rm Exc}=0$ - under this resonant condition we have achieved polariton condensation with an order of magnitude lower pumping threshold, in our previous experimental work~\cite{zasedatelev2021single}.


\subsection{Vibrational condensation}

The resonant condition discussed above is equivalent to the blue-detuned optomechanical configuration ($\Delta\Omega_{OM}=+\omega_{\rm Vib}$) outlined in Section~\ref{sec:susceptibility} which corresponds to the laser drive aligned in resonance with respect to polariton states at large wave vector ${\bf k}_{\rm ex}$ as shown in Fig.~\ref{fig:dispersion}. This blue-detuned interaction introduces anti-damping to the vibrational mode, resulting in phonon amplification as depicted in Fig.~\ref{fig:susceptibility_1D_to_2D}. Under the resonant pumping conditions, one can accumulate a significant population in this vibrational mode, transitioning into the nonlinear regime, sympathetically following the polariton occupation at the ground state. In Ref.~\cite{shishkov2023mapping}, we demonstrate the macroscopic occupation of the bright vibrational state when the system is pumped above the polariton condensation threshold $P_{th}$.

\begin{figure}
\includegraphics[width=1\linewidth]{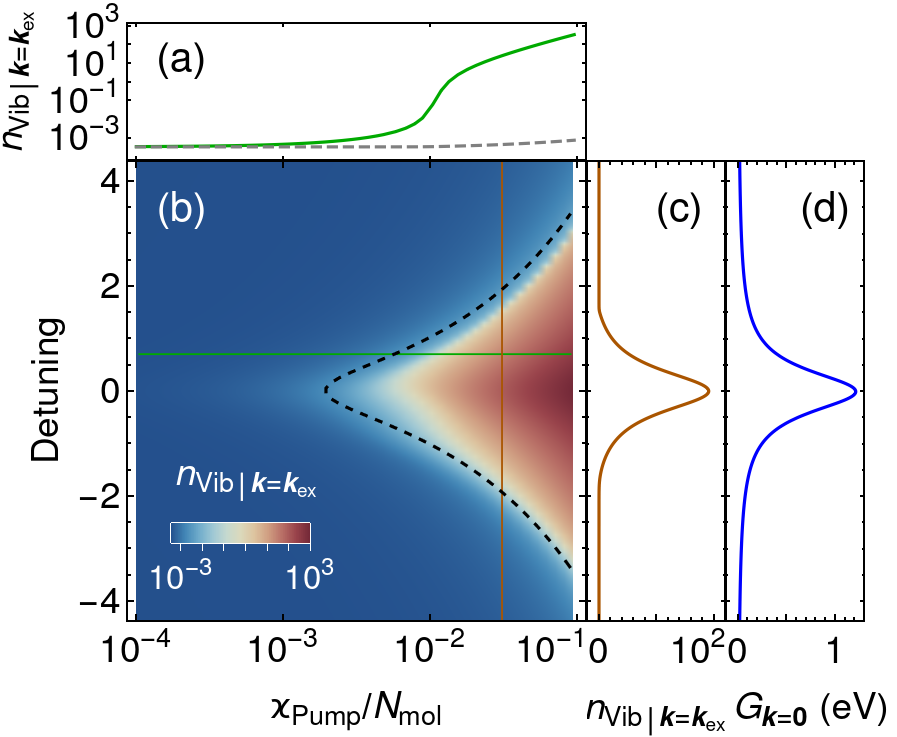}
\caption{
Vibrational occupation of the bright state with wave vector ${\bf k}_{\rm ex}$ as a function of pumping parameter -- (a), taken at the particular value of the normalized detuning $(\omega_{{\rm Pol}|{\bf k}={\bf 0}}+\omega_{\rm Vib}-\omega_{\rm Exc})/\Gamma_{\rm Exc}$ (green line) from the density plot -- (b). The black dashed line shows the analytical threshold value derived in Ref.~\cite{shishkov2023mapping}. The gray dashed line in part (a) corresponds to the occupation of dark vibrational states per molecule. The vibrational occupation as a function of the detuning at the fixed pumping conditions (brown line) -- (c). The dependence of the optomechanical constant $G_{{\bf k}={\bf 0}}$~(Eq.~(\ref{optomechanical_constant})) as a function of the normalized detuning. The parameters are the same, see Table~II.
}
    \label{fig:av_vib_occ_num}
\end{figure}

Above the threshold, polaritons undergo a stimulated thermalization process, resulting in a Bose–Einstein distribution in both momentum and energy representations~\cite{shishkov2022exact}. Similarly, due to optomechanical interaction, the vibrational degree of freedom experiences a stimulated cooling effect within their dispersion relation, leading to nonequilibrium vibrational BEC~\cite{shishkov2023mapping}. Our investigation of this transition, as a function of detuning and pumping parameters, supports the sympathetic mechanism of vibrational condensation. Fig.~\ref{fig:av_vib_occ_num} shows the results of numerical simulations for the average number of molecular vibrations. Indeed, the threshold, manifested by the superlinear increase in vibrational occupation, coincides with polariton condensation and is weakly dependent on the initial damping rate of the vibrational mode~\cite{shishkov2023mapping}. Bright molecular vibrations, akin to polaritons (Fig.~\ref{fig:av_pol_occ_num}), exhibit threshold behavior (Fig.~\ref{fig:av_vib_occ_num}a) at the same pumping rate $\varkappa_{\rm Pump}$. The change in the number of dark molecular vibrations per molecule above the condensation threshold and their contribution remains relatively small, at the level of thermal occupation $\sim10^{-3}$ at room temperature. Fig.~\ref{fig:av_vib_occ_num}c,d highlight the central role of the resonant condition, where the transition from a thermal vibrational state to the condensate appears at the onset of polariton BEC around the resonance of the blue-detuned optomechanical interaction.

\subsection{Vibrational control over polariton BEC}

All-optical control over the polariton condensation enables building ultra-fast transistor devices~\cite{zasedatelev2019room} and logic gates~\cite{baranikov2020all, sannikov2024room} with extreme low switching energy down to single photon level~\cite{zasedatelev2021single}, and compatible with silicon photonics technologies~\cite{tassan2024integrated}. However, the existing control methods necessitate signals to be resonant with the energy of polariton condensates to efficiently steer between the logic levels. This imposes stringent conditions on light sources and architecture of polariton devices, mainly revolving around 400--500 nm for the practical devices~\cite{zasedatelev2019room,baranikov2020all,sannikov2024room,tassan2024integrated}. In this work, we introduce an entirely different approach based on the optomechanical interaction through vibrational control over polariton BEC. We propose the use of non-resonant coherent anti-Stokes Raman scattering (CARS) to seed the desired vibrational mode coupled to the polariton BEC. The vibrational control offers a powerful way to manipulate the condensate, taking full advantage of the wide range of ultrafast and broadband coherent Raman microscopy methods that have been developed in recent years~\cite{ freudiger2014stimulated,polli2018broadband}. Importantly, it makes our optomechanical approach inherently compatible with fiber laser telecommunication technologies~\cite{krauss2009compact,  xu2013recent,brida2014ultrabroadband}.

We extend our analysis provided in Section~\ref{sec:susceptibility} to the polariton degrees of freedom assuming small perturbation introduced in the vibrational degree of freedom. 
The interaction part of the Hamiltonian Eq.~(\ref{H_bright}) reads $g_{{\bf k}_{\rm ex}{\bf k}} \hat s_{{\rm Low}|{\bf k}_{\rm ex}}^{\dag} \hat s_{{\rm Low}|{\bf k}} \hat C_{{\rm Vib}|{\bf k}_{\rm ex}-{\bf k}}$. As it holds certain symmetry with respect to the polariton and bright vibration operators we suggest the mechanical control over polariton BEC in close analogy to the developed optical (\textit{polariton}) control over vibrational states~\cite{zasedatelev2019room,baranikov2020all}.
We derive polariton susceptibility, $\chi_{\rm Low,eff}(\omega)$, at the ground state $\bf k = 0$ when the coherent drive is applied to the bright vibrational state with the wave vector $\bf q$
\begin{equation}
\chi_{\rm Low,eff}^{-1}(\omega)
=
\chi_{\rm Low}^{-1}(\omega)
+
\Sigma_{\rm Low}(\omega),
\end{equation}
where $\chi_{\rm Low}(\omega) = (\omega - \omega_{{\rm Low}|{\bf k}} + i\gamma_{{\rm Low}|{\bf k}}/2)^{-1}$ is the susceptibility of polaritons in the absence of the optomechanical interaction and the modification of the linear response to an external force is
\begin{multline}
\Sigma_{\rm Low}(\omega) = 
\frac
{
|g_{{\bf k}{\bf k}-{\bf q}}|^2 n_{{\rm Vib}|{{\bf q}}}
}{
\omega - \omega_{\rm Vib} - \omega_{{\rm Low}|{\bf k}-{\bf q}} +
i\Gamma_{{\rm Low}|{\bf k}-{\bf q}}/2
}
+
\\
\frac
{
|g_{{\bf k}{\bf k}+{\bf q}}|^2 n_{{\rm Vib}|{{\bf q}}}
}{
\omega + \omega_{\rm Vib} - \omega_{{\rm Low}|{\bf k}+{\bf q}} +
i\Gamma_{{\rm Low}|{\bf k}+{\bf q}}/2
}.
\end{multline}
The susceptibility $\chi_{\rm Low,eff}(\omega)$ allows us to determine the optomechanical frequency shift, $\Delta \omega_{{\rm Low}|{\bf k}} = -{\rm Re}\Sigma_{\rm Low}(\omega_{{\rm Low}|{\bf k}})$, and optomechanical damping of polaritons, $\Delta \gamma_{{\rm Low}|{\bf k}} = 2{\rm Im}\Sigma_{\rm Low}(\omega_{{\rm Low}|{\bf k}})$ induced by the vibrational control.

\begin{figure}
\includegraphics[width=1\linewidth]{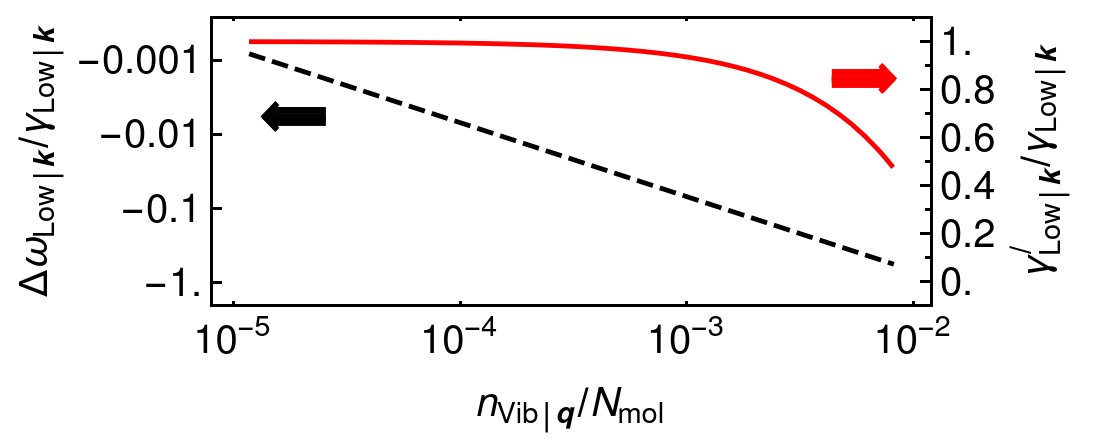}
\caption{
Frequency shift (black dashed line) and net damping of polaritons $\gamma'_{{\rm Low}{\bf k}} = \gamma_{{\rm Low}{\bf k}} + \Delta\gamma_{{\rm Low}{\bf k}}$, (red solid line)  at the ground state $k=0~{\rm \mu m}^{-1}$ when coherent drive is applied to the bright molecular vibrations with wave vector~${q}={20}~{\rm \mu m}^{-1}$.
The parameters here are $\Lambda^{2} = 1$, $\omega_{\rm Vib}=0.2$~$\rm{eV}$, $N_{\rm mol}=10^8$, $\gamma_{\rm Exc}=10^{-5}$~$\rm{eV}$, $\Gamma_{\rm Exc} = 10^{-2}$~$\rm{eV}$,
$\gamma_{\rm Vib} = 2\cdot 10^{-3}$~$\rm{eV}$, $\gamma_{\rm {Cav}|{\bf{k}}}=2.5\cdot10^{-3}$~$\rm{eV}$.
}
    \label{fig:lower_susceptibility}
\end{figure}

Fig.~\ref{fig:lower_susceptibility} demonstrates dissipative and dispersive effects created by the coherent vibrational seed. We provide both dependencies as function of vibrational quanta in the bright mode with the wave vector $\bf q$ normalized by the total number of molecules. The coherent optomechanical interaction leads to the red shift in the energy of polariton BEC as shown in Fig.~\ref{fig:lower_susceptibility}. To induce measurable red shift $|\Delta\omega_{{\rm Low} | {\bf k}}|\sim\gamma_{{\rm Low} | {\bf k}}$ one has to inject significant amount of vibrational quanta into the bright mode $n_{{\rm Vib} | {\bf q}}\geq 10^{-3} N_{\rm mol}$.

We observe similar effect in the imaginary part of the susceptibility. Fig.~\ref{fig:lower_susceptibility} shows the reduced polariton damping rate achieving $|\Delta\gamma_{{\rm Low} | {\bf k}}| \sim \gamma_{{\rm Low} | {\bf k}}$ at sufficiently large vibrational occupation $n_{{\rm Vib} | {\bf q}}\sim 10^{-2}N_{\rm mol}$. It is worth mentioning we consider steady-state problem throughout the work. Therefore, one can expect substantial improvement of the effect in dynamics. Our recent study of resonant polariton control over the BEC reveals extreme nonlinearity when the seed population injected in bursts synchronized with the onset of polariton condensation~\cite{zasedatelev2021single}. Based on the aforementioned symmetry argument in the optomechanical term of Eq.~(\ref{H_bright}) we believe non-resonant vibrational control should exhibit similar performance in dynamics, subject for further investigations.

Besides methods relying on the non-resonant coherent Raman seeding, vibrational population can be injected directly via mid-IR photons shall the vibrational mode be also dipole-allowed~\cite{chen2021continuous,xomalis2021detecting}. This offers a new interface between light-matter states in the visible and mid-IR spectral ranges that is potentially capable for single photon operation at room temperature~\cite{zasedatelev2021single}.

\section{Conclusion}

In this work, we developed an optomechanical formalism for coherent light-matter and vibrational states in exciton-polariton systems with strong exciton-phonon interactions. Our analysis reveals two distinct configurations: red- and blue-detuned optomechanical interactions. Being linearized, the blue-detuned interaction results in a \textit{two-mode squeezing} term in the Hamiltonian. This genuinely gives rise to strong correlations between vibrational and polariton states. Driven by the blue-detuned laser, this term leads to exponential growth in the occupation of both the collective vibrational mode and polaritons at the BEC. This mechanism reduces the threshold for polariton condensation, making it most efficient at the resonant condition: $\omega_{\rm Pump} = \omega_{{\rm Pol}|{\bf k}={\bf 0}}+\omega_{\rm Vib}$.

Apart from dissipative effects, vibrational degrees of freedom experience frequency changes under optomechanical interaction. The two-dimensional dispersion of exciton-polariton states, when coupled to the vibrational mode, gives rise to the effective mass of coherent molecular vibrations. Subject to laser detuning with respect to the polariton states, the effective mass can be either positive for blue-detuned interactions or negative when red-detuned. 
The effective mass acquired within the blue-detuned interaction, along with sympathetic thermalization above the threshold, allows for nonequilibrium vibrational condensation~\cite{shishkov2023mapping}.

Additionally, we propose parametric vibrational control over the polariton BEC. By utilizing coherent Raman scattering to excite vibrational modes, we introduce a non-resonant method to control the dissipative and dispersive properties of polariton states. Importantly, this new optomechanical approach circumvents the stringent resonant conditions required by existing all-optical methods, thus potentially making ultra-fast polariton logic compatible with telecommunication wavelengths.

In summary, quantum optomechanics with light-matter states not only serves as a convenient theoretical framework to describe dynamics within polariton systems but opens a new avenue with extensive implications in photochemistry, nonlinear and quantum optics. This encompasses vibrational condensation~\cite{shishkov2023mapping}, parametric control bridging visible and infrared (IR) spectral ranges, and the bipartite optomechanical~\cite{mancini1997ponderomotive, bose1997preparation, vitali2007optomechanical} and mechanical-mechanical~\cite{hartmann2008steady, liao2014entangling, teklu2018cavity} entanglement generation between excitonic and vibrational light-matter states, among many other possibilities.

\section*{Acknowledgements}

V.Yu.Sh. and E.S.A. thank Foundation for the Advancement of Theoretical Physics and Mathematics "Basis". V.Yu.Sh. and E.S.A. thank a grant from Russian Science Foundation (project No. 20-72-10057) for financial support. A.V.Z. acknowledges support from the European Union's Horizon 2020 research and innovation programme under the Marie Skłodowska-Curie grant agreement No 101030987 (LOREN). 

\newpage
\appendix

\section{Hamiltonian of a molecular system with strong light-matter and vibronic interactions} \label{appendix: Hamiltonian}

{\it Dressed vibrational and exciton states.}
Strong exciton-vibration interaction is quite common in molecular systems. 
The interplay between electronic and vibrational degrees of freedom requires transformation to the basis of dressed states.
Therefore, we substitute the initial operators $\hat \sigma_{{\rm Exc}j}$ and $\hat b_{{\rm Vib}j}$ with the operators of the dressed excitons ${\mathcal{\hat S}}_{{\rm Exc}j}$ and dressed vibrations ${\mathcal {\hat B}}_{{\rm Vib}j}$ as follows
\begin{equation} \label{dressed_excitons}
\hat \sigma_{{\rm Exc}j} 
= 
{\mathcal{\hat S}}_{{\rm Exc}j} 
e^{-\Lambda({\mathcal{\hat B}}_{{\rm Vib}j}^\dag-{\mathcal{\hat B}}_{{\rm Vib}j})},
\end{equation}
\begin{equation} \label{dressed_vibrons}
\hat b_{{\rm Vib}j} 
= 
{\mathcal{\hat B}}_{{\rm Vib}j} 
- 
\Lambda {\mathcal{\hat S}}^\dag_{{\rm Exc}j} {\mathcal{\hat S}}_{{\rm Exc}j}.
\end{equation}
This substitution preserves the commutation relations for the operators ${\mathcal{\hat S}}_{{\rm Exc}j}$ and ${\mathcal{\hat B}}_{{\rm Vib}j}$, namely the dressed operators should have the same commutator relations as the initial ones $\hat \sigma_{{\rm Exc}j}$ and $\hat b_{{\rm Vib}j}$.

The transition to the dressed state representation is equivalent to diagonalization of the initial vibrational and excitonic parts of the Hamiltonian~(\ref{Full Hamiltonian simple form}) leading to
\begin{multline}\label{FullHamiltonian_dressed_approx}
\hat H = 
\sum\limits_{\bf{k}} 
\hbar \omega_{{\rm Cav}|{\bf k}}
\hat a_{{\rm Cav}|{\bf k}}^\dag \hat a_{{\rm Cav}|{\bf k}}
+
\\
\sum_{j=1}^{N_{\rm mol}} 
\hbar \omega_{\rm Exc} 
{\mathcal{\hat S}}^\dag_{{\rm Exc}j} {\mathcal{\hat S}}_{{\rm Exc}j}
+
\sum_{j=1}^{N_{\rm mol}} 
\hbar \omega_{\rm Vib} {\mathcal{\hat B}}_{{\rm Vib}j}^\dag {\mathcal{\hat B}}_{{\rm Vib}j}
+
\\
\sum\limits_{j=1}^{N_{\rm mol}} 
\sum\limits_{{\bf k}} 
\hbar \Omega_{j{\bf k}}
\left( 
\mathcal{\hat S}_{{\rm Exc}j}^\dag \hat a_{{\rm Cav}|{\bf k}} e^{i{\bf k}{\bf r}_j}
+
h.c.
\right)
-
\\
\sum\limits_{j=1}^{N_{\rm mol}} 
\sum\limits_{{\bf k}} 
\hbar \Lambda {\Omega _{j{\bf k}}}
\left( 
\mathcal{\hat S}_{{\rm Exc}j}^\dag 
\mathcal{\hat B}_{{\rm Vib}j}
\hat a_{{\rm Cav}|{\bf k}} e^{i{\bf k}{\bf r}_j}
+
h.c.
\right)
+
\\
\sum\limits_{j=1}^{N_{\rm mol}} 
\sum\limits_{{\bf k}} 
\hbar \Lambda {\Omega _{j{\bf k}}}
\left( 
\mathcal{\hat S}_{{\rm Exc}j}^\dag 
\mathcal{\hat B}_{{\rm Vib}j}^\dag
\hat a_{{\rm Cav}|{\bf k}} e^{i{\bf k}{\bf r}_j}
+
h.c.
\right),
\end{multline} 
where we introduce the energy of the dressed exciton states $\omega_{\rm Exc} = \omega_{\rm exc} - \Lambda^2 \omega_{\rm Vib}$.
In the Hamiltonian~(\ref{FullHamiltonian_dressed_approx}) we decomposed the operator $ e^{-\Lambda({\mathcal{\hat B}}_{{\rm Vib}j}^\dag-{\mathcal{\hat B}}_{{\rm Vib}j})} \approx 1 - \Lambda({\mathcal{\hat B}}_{{\rm Vib}j}^\dag-{\mathcal{\hat B}}_{{\rm Vib}j}) $.
Thus, in the Hamiltonian~(\ref{FullHamiltonian_dressed_approx}), we preserve terms with coefficients up to $\Lambda^2 n_{\rm Vib\_per\_mol}$, where $n_{\rm Vib\_per\_mol}$ is the average occupation of the vibrational states per one molecule.
After we derive the main equations, we verify that $\Lambda^2 n_{\rm Vib\_per\_mol} \ll 1$ is always fulfilled for the system described in the main text.

{\it Bright and dark states.}
In the Hamiltonian~(\ref{FullHamiltonian_dressed_approx}) the wavevectors $\bf k$ belong to states of the electromagnetic field in the cavity. 
In most experimental systems, the number of states in an optical cavity is much smaller than the number of molecules.
Thus, a Fabry–Pérot optical cavity with the fundamental mode characterized by a wavelength $\lambda$ has the number of states equal to $N_{\rm states} \approx \pi k_{\rm max}^2S/(2 \pi)^2$, where $S$ is the area of interest (typically active or illuminated area).
In state-of-the-art BEC experiments~\cite{zasedatelev2021single, zasedatelev2019room} these parameters are the following: $\lambda \approx 500~{\rm \mu m}$ and $S \approx 500~{\rm \mu m}^2$, which result in $N_{\rm states} \approx 10^4$ number of states.
But, the number of molecules $N_{\rm mol}$ in the same area is $\approx 10^8$, therefore $N_{\rm states} \ll N_{\rm mol}$.

This analysis shows that most exciton states do not couple to the cavity and do not form of the polariton states.
The excitons that couple with the cavity are the bright excitons, while the rest states are the dark excitons.  
Bright excitons are phase-coherent, many-body delocalized states with a well-defined in-plane momentum $\hbar\bf k$, matching the corresponding eigenstates of the cavity. In contrast, the dark excitons, lacking well-defined momentum, represent a manifold of localized states. Analogous to excitons — with the sole distinction being their lack of dipole coupling to the cavity (we assume Raman active molecules) — we separate all molecular vibrations into bright or dark. Bright vibrations represent coherent and delocalized states. Similarly to bright excitons, they have a well-defined momentum $\hbar\bf k_{\rm Vib}$, unlike the localized dark vibrational states.

It is convenient to introduce collective operators that effectively describes all the bright excitons and the bright vibrations with the wave vector $\bf k$
\begin{equation} \label{ExcitonsFourier}
\hat c_{{\rm Exc}|{\bf k}} 
= 
\frac{1}{\Omega_R}
\sum_{j=1}^{N_{\rm mol}}  
\Omega_{j{\bf k}}
{\mathcal{\hat S}}_{{\rm Exc}j} {e^{-i{\bf k}{\bf r}_j}},
\end{equation}
\begin{equation} \label{VibronsFourier}
\hat c_{{\rm Vib}|{\bf k}} = 
\frac{1} {\sqrt{N_{\rm mol}}}
\sum_{j=1}^{N_{\rm mol}} 
{\mathcal{\hat B}}_{{\rm Vib}j} e^{-i{\bf k}{\bf r}_j},
\end{equation}
where $\Omega_R = \sqrt{\sum_j|\Omega_{j{\bf k}}|^2}$ is the Rabi frequency which we assume to be $\bf k$ independent.
The operators of bright molecular vibrations obey bosonic commutation relation $\left[ \hat c_{{\rm Vib}|{\bf k}},\hat c_{{\rm Vib}|{\bf k'}}^\dag \right] = \delta_{{\bf k},{\bf k'}}$.
In the limit of large $N_{\rm mol}$ and small enough excitation density, the operators of the bright excitons also exhibit bosonic properties $\left[ \hat c_{{\rm Exc}|{\bf k}},\hat c_{{\rm Exc}|{\bf k'}}^\dag \right] \approx \delta_{{\bf k},{\bf k'}}$~\cite{combescot2008microscopic}. 
Using operators~(\ref{ExcitonsFourier})--(\ref{VibronsFourier}) 
we transform the Hamiltonian~(\ref{FullHamiltonian_dressed_approx}) 

\begin{multline}\label{FullHamiltonian_bright}
\hat H = 
\sum\limits_{\bf k} 
\hbar \omega _{{\rm Cav}|{\bf k}}
\hat a_{{\rm Cav}|{\bf k}}^\dag \hat a_{{\rm Cav}|{\bf k}}
+
\\
\sum\limits_{\bf k} 
\hbar \omega_{\rm Exc} 
\hat c_{{\rm Exc}|{\bf k}}^\dag \hat c_{{\rm Exc}|{\bf k}}
+
\sum\limits_{\bf{k}} 
\hbar \omega_{\rm Vib}\hat c_{{\rm Vib}|{\bf k}}^\dag \hat c_{{\rm Vib}|{\bf k}}
+
\\
\sum\limits_{\bf k} 
\hbar \Omega_R
\left( 
\hat c_{{\rm Exc}|{\bf k}}^\dag \hat a_{{\rm Cav}|{\bf k}} 
+ 
\hat c_{{\rm Exc}|{\bf k}} \hat a_{{\rm Cav}|{\bf k}}^\dag 
\right)
-
\\
\sum\limits_{{\bf k}, {\bf k}'} 
\hbar \frac{\Lambda \Omega_R}{\sqrt{N_{\rm mol}}}
\left( 
\hat c_{{\rm Exc}{\bf k'}}^\dag
\hat c_{{\rm Vib}|{\bf k' }-{\bf k}}
\hat a_{{\rm Cav}|{\bf k}}
+
h.c.
\right)
+
\\
\sum\limits_{{\bf k}, {\bf k}'} 
\hbar \frac{\Lambda \Omega_R}{\sqrt{N_{\rm mol}}}
\left( 
\hat c_{{\rm Exc}{\bf k'}}^\dag
\hat c_{{\rm Vib}|{\bf k}-{\bf k'}}^\dag
\hat a_{{\rm Cav}|{\bf k}}
+
h.c.
\right)
+
\\
N_{\rm mol} \hbar \omega_{\rm Exc} \hat n_{\rm Exc_D}
+
N_{\rm mol} \hbar \omega_{\rm Vib} \hat n_{\rm Vib_D}
+
\hat H_{\rm cav-dark}.
\end{multline} 
We denote the operator of the number of the dark excitons per one molecule as $\hat n_{\rm Exc_D}$ and the similar operator for the dark vibrations as $\hat n_{\rm Vib_D}$
\begin{equation} \label{H_dark_exc}
\hat n_{\rm Exc_D}
=
\frac{1}{N_{\rm mol}}
\sum_{j=1}^{N_{\rm mol}} 
{\mathcal{\hat S}}^\dag_{{\rm Exc}j} {\mathcal{\hat S}}_{{\rm Exc}j}
-
\frac{1}{N_{\rm mol}}
\sum\limits_{\bf{k}}
\hat c_{{\rm Exc}|{\bf k}}^\dag \hat c_{{\rm Exc}|{\bf k}},
\end{equation}
\begin{equation}\label{H_dark_vib}
\hat n_{\rm Vib_D}
=
\frac{1}{N_{\rm mol}}
\sum_{j=1}^{N_{\rm mol}} 
{\mathcal{\hat B}}_{{\rm Vib}j}^\dag {\mathcal{\hat B}}_{{\rm Vib}j}
-
\frac{1}{N_{\rm mol}}
\sum\limits_{\bf k} 
\hat c_{{\rm Vib}|{\bf k}}^\dag \hat c_{{\rm Vib}|{\bf k}}.
\end{equation}
We also introduced the Hamiltonian of the indirect interactions between the dark states and cavity modes~$\hat H_{{\rm cav-dark}}$
\begin{multline} \label{H_dark_int}
\hat H_{\rm cav-dark}
= 
\\
-
\sum\limits_{j=1}^{N_{\rm mol}} 
\sum\limits_{\bf k} 
\hbar \Lambda \Omega_{j{\bf k}} 
\left( 
\mathcal{\hat S}_{{\rm Exc}j}^\dag 
\mathcal{\hat B}_{{\rm Vib}j}
\hat a_{{\rm Cav}|{\bf k}} e^{i{\bf k}{\bf r}_j}
+
h.c.
\right)
+
\\
\sum\limits_{j=1}^{N_{\rm mol}} 
\sum\limits_{\bf k} 
\hbar \Lambda \Omega_{j{\bf k}} 
\left( 
\mathcal{\hat S}_{{\rm Exc}j}^\dag 
\mathcal{\hat B}_{{\rm Vib}j}^\dag
\hat a_{{\rm Cav}|{\bf k}} e^{i{\bf k}{\bf r}_j}
+
h.c.
\right)
+
\\
\sum\limits_{{\bf k}, {\bf k}'} 
\hbar \frac{\Lambda \Omega_R}{\sqrt{N_{\rm mol}} }
\left( 
\hat c_{{\rm Exc}|{\bf k' }}^\dag
\hat c_{{\rm Vib}|{\bf k' }-{\bf k }}
\hat a_{{\rm Cav}|{\bf k }}
+
h.c.
\right)
-
\\
\sum\limits_{{\bf k}, {\bf k}'} 
\hbar \frac{\Lambda \Omega_R}{\sqrt{N_{\rm mol}} }
\left( 
\hat c_{{\rm Exc}|{\bf k' }}^\dag
\hat c_{{\rm Vib}|{\bf k }-{\bf k' }}^\dag
\hat a_{{\rm Cav}|{\bf k }} 
+
h.c.
\right)
.
\end{multline}

{\it Polariton states.}
Strong interaction between the excitons and the electromagnetic field of the cavity leads to the formation of the new collective states -- exciton-polaritons.
It follows from the Hamiltonian~(\ref{FullHamiltonian_bright}), that only bright excitons directly interact with the cavity.
Unlike bare photons in the cavity, polaritons have the material component, that allows them to thermalize towards BEC above the condensation threshold. 
We introduce operators for the lower $\hat s_{{\rm Low}|{\bf k}}$ and upper $\hat s_{{\rm Up}|{\bf k}}$ polaritons 
\begin{equation} \label{TransformationForLowerPolaritons}
\hat s_{{\rm Low}|{\bf k}} 
= 
\hat a_{{\rm Cav}|{\bf k}} \cos \varphi_{{\bf k}} 
- 
\hat c_{{\rm Exc}|{\bf k}} \sin \varphi_{{\bf k}},
\end{equation}
\begin{equation} \label{TransformationForUpperPolaritons}
\hat s_{{\rm Up}|{\bf k}} 
= 
\hat a_{{\rm Cav}|{\bf k}} \sin \varphi_{{\bf k}} 
+
\hat c_{{\rm Exc}|{\bf k}} \cos \varphi_{{\bf k}},
\end{equation}
where
\begin{multline} \label{TransformationAngle}
\varphi_{\bf k}=
{\rm arctg}
\Bigg[ 
\sqrt{
\frac
{(\omega_{\rm Exc}-\omega_{{\rm Cav}|{\bf k}})^2+4\Omega_R^2}
{4\Omega_R^2}
}
 \\ 
-
\frac{\omega_{\rm Exc}-\omega_{{\rm Cav}|{\bf k}}}{2\Omega_R}
\Bigg]
\end{multline}
The dispersion curves of the lower and the upper polariton states are
\begin{equation} \label{FrequenciesOfLowerPolaritons}
\omega _{{\rm Low}|{\bf k}} 
= 
\frac{\omega_{\rm Exc} + \omega_{{\rm Cav}|{\bf k}} }{ 2}
-
\sqrt{
\frac{
\left( 
\omega_{\rm Exc} - \omega _{{\rm Cav}|{\bf k}} 
\right)^2 
}{
4
}
+
\Omega_R^2
},
\end{equation}
\begin{equation} \label{FrequenciesOfUpperPolaritons}
\omega _{{\rm Up}|{\bf k}} 
= 
\frac{\omega_{\rm Exc} + \omega_{{\rm Cav}|{\bf k}} }{ 2}
+
\sqrt{
\frac{
\left( 
\omega_{\rm Exc} - \omega _{{\rm Cav}|{\bf k}} 
\right)^2 
}{ 4
}
+
\Omega_R^2
}.
\end{equation}

Thus, we obtain~(\ref{FullHamiltonian_polaritons}).

\section{Interaction with the environment: dissipation, pumping, relaxation and thermalization} \label{appendix: Relaxation}
A density matrix $\hat \rho$ characterizes the current state of the system.
The density matrix is governed by the master equation~\cite{scully1997quantum, carmichael2009open}
\begin{equation} \label{MasterEquation}
{\frac{ d\hat \rho}{dt}} 
= 
{\frac{ i}{\hbar}}{\left[ {\hat \rho}, {\hat H} \right]}
+
\sum_n L_n(\hat \rho),
\end{equation}
where $L_n(\hat \rho)$ is the Lindblad superoperator 
\begin{equation}
L_n(\hat \rho) 
=
\gamma_n
\left(
\hat A_n \hat \rho \hat A_n^\dag 
- 
\frac{1 }{ 2} \hat \rho \hat A_n^\dag \hat A_n
- 
\frac{1 }{ 2} \hat A_n^\dag \hat A_n \hat \rho 
\right),
\end{equation}
$\hat A_n$ is the relaxation operator, and $\gamma_n$ is the relaxation rate.
The Lindblad superoperators can describe both the dissipation and pumping of the system.

Below, we consider the relaxation processes in each of the subsystems separately.


The energy relaxation rate of dressed exciton states in the absence of a cavity is defined by $\gamma_{\rm Exc}$.
The corresponding relaxation operator for $j$-th molecule is $\hat {\mathcal{S}}_{{\rm Exc}j}$.
The dressed exciton states also undergo dephasing processes at the rate $\Gamma_{\rm Exc}$ that corresponds to the relaxation operator $\hat {\mathcal{S}}_{{\rm Exc}j}^\dag \hat {\mathcal{S}}_{{\rm Exc}j}$.
Usually, in molecular systems at room temperature the exciton energy dissipation rate $\gamma_{\rm Exc}$ is much smaller than the dephasing $\Gamma_{\rm Exc}$.
The corresponding Lindblad superoperators are
\begin{multline} \label{L_exc_diss}
L_{\rm Exc(diss)} = 
\sum_{j=1}^{N_{\rm mol}}
\frac{\gamma_{\rm Exc}}{2}
\left(
2\hat {\mathcal{S}}_{{\rm Exc}j} \hat \rho \hat {\mathcal{S}}_{{\rm Exc}j}^\dag
-
\right.
\\
\left.
\hat {\mathcal{S}}_{{\rm Exc}j}^\dag \hat {\mathcal{S}}_{{\rm Exc}j} \hat \rho
-
\hat \rho \hat {\mathcal{S}}_{{\rm Exc}j}^\dag \hat {\mathcal{S}}_{{\rm Exc}j} 
\right),
\end{multline}
\begin{multline} \label{L_exc_deph}
L_{\rm Exc(deph)} = 
\sum_{j=1}^{N_{\rm mol}}
\frac{\Gamma_{\rm Exc}}{2}
\left(
2\hat {\mathcal{S}}_{{\rm Exc}j}^\dag \hat {\mathcal{S}}_{{\rm Exc}j} \hat \rho \hat {\mathcal{S}}_{{\rm Exc}j}^\dag \hat {\mathcal{S}}_{{\rm Exc}j}
-
\right.
\\
\left.
\hat {\mathcal{S}}_{{\rm Exc}j}^\dag \hat {\mathcal{S}}_{{\rm Exc}j} \hat \rho
-
\hat \rho \hat {\mathcal{S}}_{{\rm Exc}j}^\dag \hat {\mathcal{S}}_{{\rm Exc}j} 
\right).
\end{multline}

The dressed vibrational states dissipate their energy with the rate $\gamma_{\rm Vib}$.
Since thermal fluctuations affect the system at room temperature, the corresponding relaxation operators of dressed vibrational states for the $j$-th molecule are $\mathcal{\hat B}_{{\rm Vib}j}$ and $\mathcal{\hat B}_{{\rm Vib}j}^\dag$ with the relaxation rates $(1 + n_{\rm Vib}^{\rm th})\gamma_{\rm Vib}$ and $n_{\rm Vib}^{\rm th}\gamma_{\rm Vib}$, where $n_{\rm Vib}^{\rm th} = 1/({\rm exp}(\hbar \omega_{\rm Vib}/k_BT)-1)$ is the mean of the thermal distribution of molecular vibrations at the temperature $T$.
The corresponding Lindblad superoperators are
\begin{multline} \label{L_vib}
L_{\rm Vib} = 
\sum_{j=1}^{N_{\rm mol}}
\frac{\gamma_{\rm Vib}(1 + n_{\rm Vib}^{\rm th})}{2}
\left(
2\hat {\mathcal{B}}_{{\rm Vib}j} \hat \rho \hat {\mathcal{B}}_{{\rm Vib}j}^\dag
-
\right.
\\
\left.
\hat {\mathcal{B}}_{{\rm Vib}j}^\dag \hat {\mathcal{B}}_{{\rm Vib}j} \hat \rho
-
\hat \rho \hat {\mathcal{B}}_{{\rm Vib}j}^\dag \hat {\mathcal{B}}_{{\rm Vib}j} 
\right)
+
\\
\sum_{j=1}^{N_{\rm mol}}
\frac{\gamma_{\rm Vib}n_{\rm Vib}^{\rm th}}{2}
\left(
2\hat {\mathcal{B}}_{{\rm Vib}j}^\dag \hat \rho \hat {\mathcal{B}}_{{\rm Vib}j}
-
\hat {\mathcal{B}}_{{\rm Vib}j} \hat {\mathcal{B}}_{{\rm Vib}j}^\dag \hat \rho
-
\right.
\\
\left.
\hat \rho \hat {\mathcal{B}}_{{\rm Vib}j} \hat {\mathcal{B}}_{{\rm Vib}j}^\dag 
\right).
\end{multline}

The dissipation rate remains the same $\gamma_{\rm Vib}$ for both dark and bright vibrational states.
The same applies to dark excitons.
The energy dissipation rate of the dark excitons equals $\gamma_{\rm Exc}$.
However, due to the strong interaction between the cavity and bright exciton states, the dissipation rates of the polariton states are different.
The dissipation rates of the lower polariton states $\gamma_{{\rm Low}|{\bf k}}$ and for upper polariton states $\gamma_{{\rm Up}|{\bf k}}$ are determined by the Hopfield coefficients and equal $\gamma_{{\rm Low}|{\bf k}} = \gamma_{{\rm Cav}|{\bf k}} \cos^2\varphi_{\bf k} + \gamma_{{\rm Exc}} \sin^2\varphi_{\bf k}$ and $\gamma_{{\rm Up}|{\bf k}} = \gamma_{\rm Exc} \cos^2\varphi_{\bf k} + \gamma_{{\rm Cav}|{\bf k}} \sin^2\varphi_{\bf k}$, where $\gamma_{{\rm Cav}|{\bf k}}$ is the dissipation rate of cavity photons with the wave vector $\bf k$.
The corresponding relaxation operators are $\hat s_{{\rm Low}|{\bf k}}$ and $\hat s_{{\rm Up}|{\bf k}}$.
The corresponding Lindblad superoperators are
\begin{multline} \label{L_up}
L_{\rm Low} = 
\sum_{\bf k}
\frac{\gamma_{{\rm Low}|{\bf k}}}{2}
\left(
2\hat s_{{\rm Low}|{\bf k}} \hat \rho \hat s_{{\rm Low}|{\bf k}}^\dag
-
\right.
\\
\left.
\hat s_{{\rm Low}|{\bf k}}^\dag \hat s_{{\rm Low}|{\bf k}} \hat \rho
-
\hat \rho \hat \hat s_{{\rm Low}|{\bf k}}^\dag \hat s_{{\rm Low}|{\bf k}} 
\right)
\end{multline}
\begin{multline} \label{L_low}
L_{\rm Vib} = 
\sum_{\bf k}
\frac{\gamma_{{\rm Up}|{\bf k}}}{2}
\left(
2\hat s_{{\rm Up}|{\bf k}} \hat \rho \hat s_{{\rm Up}|{\bf k}}^\dag
-
\right.
\\
\left.
\hat s_{{\rm Up}|{\bf k}}^\dag \hat s_{{\rm Up}|{\bf k}} \hat \rho
-
\hat \rho \hat s_{{\rm Up}|{\bf k}}^\dag \hat s_{{\rm Up}|{\bf k}} 
\right)
\end{multline}

Being hybrid light–matter states, polaritons inherit properties from both the molecules and the electromagnetic field of the cavity.  
Recent a microscopic theory suggested that polariton thermalization originates from the low-energy molecular vibrations coupled to the material component of the polaritons~\cite{tereshchenkov2024thermalization} and thermalization processes between lower polariton states with the wavevector ${\bf k}_1$ and ${\bf k}_2$ can be described by the relaxation operators ${{\hat s}_{{\rm Low}|{{\bf k}_2}}}\hat s_{{\rm Low}|{{\bf k}_1}}^\dag$, ${{\hat s}_{{\rm Low}|{{\bf k}_1}}}\hat s_{{\rm Low}|{{\bf k}_2}}^\dag$ and  thermalization rates $\gamma _{\rm therm}^{{{\bf k}_2}{{\bf k}_1}}$ and $\gamma _{\rm therm}^{{{\bf k}_1}{{\bf k}_2}}$. 
The corresponding Lindblad superoperators are
\begin{multline} \label{Lindblad for polaritons}
L_{\rm therm}(\hat \rho) =
\sum_{{\bf k}_1, {\bf k}_2} 
\frac{\gamma_{\rm therm}^{{\bf k}_1{\bf k}_2}}{2} 
\left(
2\hat s_{{\rm low}{\bf k }_2} \hat s_{{\rm low}{\bf k }_1}^\dag 
\hat \rho
\hat s_{{\rm low}{\bf k }_1} \hat s_{{\rm low}{\bf k }_2}^\dag
-
\right.
\\
\hat s_{{\rm low}{\bf k }_1} \hat s_{{\rm low}{\bf k }_2}^\dag
\hat s_{{\rm low}{\bf k }_2} \hat s_{{\rm low}{\bf k }_1}^\dag 
\hat \rho
-
\\
\left.
\hat \rho
\hat s_{{\rm low}{\bf k }_1} \hat s_{{\rm low}{\bf k }_2}^\dag
\hat s_{{\rm low}{\bf k }_2} \hat s_{{\rm low}{\bf k }_1}^\dag 
\right)
\end{multline}
The ratio between the rates upward and downward thermalization processes $\gamma_{{\rm therm}}^{{{\bf k}_1}{{\bf k}_2}}$ and $\gamma_{{\rm therm}}^{{{\bf k}_2}{{\bf k}_1}}$ is determined by the Kubo--Martin--Schwinger relation~\cite{kubo1957statistical}
\begin{equation} \label{kubo-martin-schwinger relation}
\gamma_{\rm therm}^{{\bf k}_1{\bf k}_2} 
= 
\gamma_{\rm therm}^{{\bf k}_2{\bf k}_1} 
\exp 
\left( 
\frac{ \hbar\omega_{{\rm low}{\bf k}_2} - \hbar\omega_{{\rm low}{\bf k}_1} }{ k_BT } 
\right)
.
\end{equation}
where $T$ is the temperature of the environment.

We consider the resonant pumping scheme where the laser excites bright excitonic states at high wavevector ${\bf k}={\bf k}_{\rm ex}$ (the lower and upper polaritons).
This process can be described by the relaxation operators 
$ \hat s_{{\rm Low}|{\bf k}_{\rm ex}}$, $\hat s_{{\rm Low}|{\bf k}_{\rm ex}}^\dag$, $ \hat s_{{\rm Up}|{\bf k}_{\rm ex}}$ and $ \hat s_{{\rm Up}|{\bf k}_{\rm ex}}^\dag$ and corresponding rates $\varkappa_{{\rm Low}|{\bf k}_{\rm ex}}$, $\varkappa_{{\rm Low}|{\bf k}_{\rm ex}}$, $\varkappa_{{\rm Up}|{\bf k}_{\rm ex}}$ and $\varkappa_{{\rm Up}|{\bf k}_{\rm ex}}$~\cite{shishkov2022exact, shishkov2022analytical}.
We distinguish between the lower polaritons having relatively small in-plane momenta within $kT \simeq 30~{\rm meV}$ range of energy around the ground state ${\hbar\bf k}={\bf 0}$  and the polariton states with the large momentum ${\hbar\bf k}_{\rm ex}$ as represented by the dispersion curves in ~(Fig.~\ref{fig:dispersion}).
The former we call ``polaritons'' and the latter we call ``bright excitons''.
The reason for this separation is that lower polaritons with small wave vectors have a dominant electromagnetic component and are not pumped directly, whereas the polaritons with large wave vectors have a dominant exciton component and undergo direct optical pumping.
For large wave vector ${\bf k}_{\rm ex}$ we do not consider upper polariton state, because we assume that $\omega_{{\rm Up}|{\bf k}_{\rm ex}} \approx \omega_{{\rm Cav}|{\bf k}_{\rm ex}}$, $\omega_{{\rm Low}|{\bf k}_{\rm ex}} \approx \omega_{\rm Exc}$, and the optical pumping is aligned in resonance with the exciton energy as shown in Fig.~\ref{fig:dispersion}.
Therefore, we can omit the dynamics of the upper polaritons with the wavevector ${\bf k}_{\rm ex}$ and denote  
\begin{equation}
\varkappa_{{\rm Low}|{\bf k}_{\rm ex}} = \varkappa_{\rm Pump}  
\end{equation}
and set
\begin{equation}
\hat s_{{\rm Low}|{\bf k}_{\rm ex}} \approx \hat c_{{\rm Exc}|{\bf k}_{\rm ex}}.
\end{equation}
Thus, we obtain the Lindblad suparoperators for the resonant pumping
\begin{multline} \label{L_vib}
L_{\rm Pump} = 
\frac{\varkappa_{\rm Pump}}{2}
\left(
2\hat c_{{\rm Exc}|{\bf k}_{\rm ex}} \hat \rho \hat c_{{\rm Exc}|{\bf k}_{\rm ex}}^\dag
-
\right.
\\
\left.
\hat c_{{\rm Exc}|{\bf k}_{\rm ex}}^\dag \hat c_{{\rm Exc}|{\bf k}_{\rm ex}} \hat \rho
-
\hat \rho \hat c_{{\rm Exc}|{\bf k}_{\rm ex}}^\dag \hat c_{{\rm Exc}|{\bf k}_{\rm ex}} 
\right)
+
\\
\frac{\varkappa_{\rm Pump}}{2}
\left(
2\hat c_{{\rm Exc}|{\bf k}_{\rm ex}}^\dag \hat \rho \hat c_{{\rm Exc}|{\bf k}_{\rm ex}}
-
\hat c_{{\rm Exc}|{\bf k}_{\rm ex}} \hat c_{{\rm Exc}|{\bf k}_{\rm ex}}^\dag \hat \rho
-
\right.
\\
\left.
\hat \rho \hat c_{{\rm Exc}|{\bf k}_{\rm ex}} \hat c_{{\rm Exc}|{\bf k}_{\rm ex}}^\dag 
\right).
\end{multline}

The lower polariton states with large wavevectors have a dominant excitonic nature, unlike upper polaritons with the same momentum. Practically, they degenerate in energy with the localized, dark excitons. The total number of dark exciton states significantly outnumbers that of the polariton states. As a result, the dark excitons effectively act as a reservoir for the polariton states with large wavevectors, especially for the lower polariton state with a given wavevector ${\bf k}_{\rm ex}$.
We phenomenologically account for this mechanism, assuming that the corresponding relaxation rate is $\gamma_{\rm Exc}^{\rm B-D} = \Gamma_{\rm Exc}$.
A similar principle applies to the bright and dark vibrational states.
We set the corresponding relaxation rate $\gamma_{\rm Vib}^{\rm B-D}$ to $\gamma_{\rm Vib}/2$.

\section{Discretization of the system} \label{appendix: Discretization}
To provide a numerical simulation, we discretize the system and map its momenta states onto the energy space, moving from the continuum of wavevectors to a discrete set of frequencies. 
We consider a discrete finite set of frequencies (${\omega _0}$, ${\omega _1}$, ${\omega _2}$, ..., ${\omega _N}$) with a sampling rate equal to $\delta \omega $, such that ${\omega _{j + 1}} - {\omega _j} = \delta \omega $. 
The total number of frequencies that we account for is $N+1$.
For the sake of simplicity, we introduce the ground state frequency as ${\omega _0} = {\omega_{{\rm Pol}|{\bf k} = {\bf 0}} }$.
Likewise, we introduce $\varphi_j=\varphi_{\bf k}|_{\omega_{{\rm Pol}|{\bf k}}=\omega_j}$
We denote the frequency intervals $R_j=(\omega_j-\delta\omega, \omega_j]$ and introduce occupation numbers and parameters of the discretized system
\begin{equation}
n_{{\rm Pol}0} = n_{{\rm Pol}|{\bf k} = {\bf 0}}
,
\end{equation}
\begin{equation}
n_{{\rm Pol}j} = \sum_{ {\bf k}: \; \omega_{{\rm Pol}|{\bf k}} \in R_j} n_{{\rm Pol}|{\bf k}}
,
\;\;
1<j<N
,
\end{equation}
\begin{equation}
n_{{\rm Vib}0} = n_{{\rm Vib}|{\bf k}={\bf 0}}
,
\end{equation}
\begin{equation}
n_{{\rm Vib}j} = \sum_{ {\bf k}: \; \omega_{{\rm Pol}|{\bf k}} \in R_j} n_{{\rm Vib}|{\bf k_{\rm ex}}-{\bf k}}
,
\;\;
1<j<N
,
\end{equation}
\begin{equation}
\gamma_{{\rm Pol}j} = \,\left. \gamma_{{\rm Pol}|{\bf k}} \right|_{\omega_{{\rm Pol}|{\bf k}} = \omega_j}
,
\end{equation}
\begin{equation}
\gamma_{\rm therm}^{jm} = \left. \gamma_{\rm therm}^{\bf k'k''} \right|_{\omega_{{\rm Pol}|{\bf k'}} = \omega _j,\,\omega _{{\rm Pol}|{\bf k''}} = \omega_m}
,
\end{equation}
\begin{equation} 
G_{j}
=
\frac{ \Lambda^2 \Omega_R^2  \Gamma_{\rm Exc} \cos^2\varphi_j }{ (\omega_{\rm Exc}-\omega_j-\omega_{\rm Vib})^2 + (\Gamma_{\rm Exc}/2)^2 }
,
\end{equation}
\begin{equation} 
\tilde G_j
=
G_j \sin^2 \varphi_{{\bf k}_{\rm ex}}.
\end{equation}
The total number of the lower polariton states within the frequency interval $\delta \omega$ close to $\omega_j$ we denote as $D_j$ and define as follows
\begin{equation} \label{NumberOfState0}
D_0 = 1 ,
\end{equation}
\begin{equation} \label{NumberOfStates}
D_j = \sum_{ {\bf k}: \; \omega_{{\rm Pol}|{\bf k}} \in R_j} 1, \;\; 1<j<N. 
\end{equation}
One can show that $D_j = S \delta\omega/(4 \pi \alpha_{\rm Pol})$ for $1<j<N$ where $S$ is the pumped area.

As a result we obtain the following discrete version of  Eq.~(\ref{NumberOfBrightExcitons})--(\ref{NumberOfBrightmolecular vibrations}) 
\begin{multline} \label{NumberOfBrightExcitonsDiscrete}
\frac{d n_{{\rm Exc}|{\bf k}_{\rm ex}}}{dt} 
=  
-  \gamma_{\rm Exc} ( n_{{\rm Exc}|{\bf k}_{\rm ex}} - \varkappa_{\rm pump} )
+ \gamma_{\rm Exc}^{\rm B-D} ( n_{\rm Exc_D} - \\ n_{{\rm Exc}|{\bf k}_{\rm ex}} ) 
- 
\sum_{m=0}^N 
\frac{\tilde G_{m}}{N_{\rm mol}} 
\left[
n_{{\rm Exc}|{\bf k}_{\rm ex}} \left(n_{{\rm Pol}m} + D_m \right) 
+
\right. \\ \left.
n_{{\rm Vib}m} \left( n_{{\rm Exc}|{\bf k}_{\rm ex}} - \frac{ n_{{\rm Pol}m} }{  D_m } \right) 
\right],
\end{multline}

\begin{multline} \label{NumberOfDarkExcitonsDiscrete}
\frac{d n_{\rm Exc_D}}{dt} 
=  
-  \gamma_{\rm Exc} n_{\rm Exc_D}
+ \frac{ \gamma_{\rm Exc}^{\rm B-D} }{ N_{\rm mol}} ( n_{{\rm Exc}|{\bf k}_{\rm ex}} - n_{\rm Exc_D} ) 
-
\\
\sum_{m=0}^N 
\frac{ G_m}{N_{\rm mol}} 
\left[
n_{\rm Exc_D} \left(n_m + D_m \right) 
+
n_{\rm Vib_D} \left( D_m n_{\rm Exc_D} - n_m \right) 
\right],
\end{multline}

\begin{multline} \label{NumberOfLowerPolaritonsHDiscrete}
\frac{d n_{{\rm Pol}j}}{dt} 
=  
-  \gamma_{{\rm Pol}j} n_{{\rm Pol}j}
+ 
G_j
\left[
n_{\rm Exc_D} \left(n_{{\rm Pol}j} + D_j \right) 
+
\right. \\ \left.
n_{\rm Vib_D} \left( D_j n_{\rm Exc_D} - n_{{\rm Pol}j} \right) 
\right]
+ 
\frac{G_j }{ N_{\rm mol} } 
\left[
n_{{\rm Exc}|{\bf k}_{\rm ex}} \left(n_{{\rm Pol}j} + D_j \right) 
+
\right. \\ \left.
n_{{\rm Vib}j} \left( n_{{\rm Exc}|{\bf k}_{\rm ex}} - \frac{ n_{{\rm Pol}j} }{ D_j} \right) 
\right]
+
\sum_{m=0}^N 
\left\{ \gamma _{\rm therm}^{{jm}}
\right. \\ \left.
\left( n_{{\rm Pol}j} + D_j \right){n_{{\rm Pol}m}} 
 - 
\gamma _{{\rm therm}}^{mj}\left( n_{{\rm Pol}m} + D_m \right) n_{{\rm Pol}j}  \right\} ,
\end{multline}

\begin{multline} \label{NumberOfDarkVibronsDiscrete}
\frac{d n_{\rm Vib_D}}{dt} 
=  
-  \gamma_{\rm Vib} \left( n_{\rm Vib_D}- n_{\rm Vib}^{\rm th}  \right) 
+ \frac{\gamma_{\rm Vib}^{\rm B-D} }{ N_{\rm mol}} \sum_{m=0}^N 
( n_{{\rm Vib}m}
\\
- D_m n_{\rm Vib_D} )
+ 
\sum_{m=0}^N 
\frac{ G_m }{ N_{\rm mol}}  
\left[
n_{\rm Exc_D} \left(n_{{\rm Pol}m} + D_m \right)
\right. \\ \left.
+
n_{\rm Vib_D} \left( D_m n_{\rm Exc_D} - n_{{\rm Pol}m} \right) 
\right],
\end{multline}
\begin{multline} \label{NumberOfBrightVibronsDiscrete}
\frac{d n_{{\rm Vib}j}}{dt} 
=  
-  \gamma_{\rm Vib} \left( n_{{\rm Vib}j} - D_j n_{\rm Vib}^{\rm th}  \right) 
+ \gamma_{\rm Vib}^{\rm B-D} 
( D_j n_{\rm Vib_D} 
\\
- n_{{\rm Vib}j} )
+
\frac{G_j}{N_{\rm mol}}  
\left[
n_{{\rm Exc}|{\bf k}_{\rm ex}} \left(n_{{\rm Pol}j} + D_j \right) 
+
\right. \\ \left.
n_{{\rm Vib}j} \left( n_{{\rm Exc}|{\bf k}_{\rm ex}} - \frac{ n_{{\rm Pol}j} }{ D_j } \right) 
\right],
\end{multline}
where $j = 0, 1, ..., N$.
The total amount of the equations is $2N+7$.
For our simulations, we set $N=100$ that fill the bottom of lower polariton dispersion within the momenta range of [$|k_{\rm min}|=0$~$\mu {\rm m}^{-1}, |k_{\rm max}|=2$~$\mu {\rm m}^{-1}$]~(Fig.~\ref{fig:dispersion}).


\bibliography{main-vPRLwRefs}

\end{document}